\documentclass[prx, twocolumn,amsmath,10pt,amssymb,aps,superscriptaddress,showkeys,showpacs]{revtex4-2}

\usepackage[utf8]{inputenc}

\usepackage{lipsum}
\usepackage{tikz}  
\usepackage{enumitem,mathtools}
\usepackage{graphicx}
\usepackage{dcolumn}
\usepackage{array}
\usepackage{bm}
\usepackage{hyperref}

\begin{document}

\title{Inferring genotype-phenotype maps using attention models}

\author{Krishna Rijal}
\affiliation{Department of Physics, Boston University, Boston, MA}

\author{Caroline M. Holmes}
\affiliation{Department of Organismic and Evolutionary Biology, Harvard University, Cambridge, MA}

\author{Samantha Petti}
\affiliation{Department of Mathematics, Tufts University, Medford, MA}

\author{Gautam Reddy}
\affiliation{Department of Physics, Princeton University, Princeton, NJ}

\author{Michael M. Desai}
\affiliation{Department of Organismic and Evolutionary Biology, Harvard University, Cambridge, MA}

\author{Pankaj Mehta}
\affiliation{Department of Physics, Boston University, Boston, MA}

\begin{abstract}
Predicting phenotype from genotype is a central challenge in genetics. Traditional approaches in quantitative genetics typically analyze this problem using methods based on linear regression. These methods generally assume that the genetic architecture of complex traits can be parameterized in terms of an additive model, where the effects of loci are independent, plus (in some cases) pairwise epistatic interactions between loci. However, these models struggle to analyze more complex patterns of epistasis or subtle gene-environment interactions. Recent advances in machine learning, particularly attention-based models, offer a promising alternative. Initially developed for natural language processing, attention-based models excel at capturing context-dependent interactions and have shown exceptional performance in predicting protein structure and function. Here, we apply attention-based models to quantitative genetics. We analyze the performance of this attention-based approach in predicting phenotype from genotype using simulated data across a range of models with increasing epistatic complexity, and using experimental data from a recent quantitative trait locus mapping study in budding yeast. We find that our model demonstrates superior out-of-sample predictions in epistatic regimes compared to standard methods. We also explore a more general multi-environment attention-based model to jointly analyze  genotype-phenotype  maps across multiple environments and show that such architectures can be used for ``transfer learning'' -- predicting phenotypes in novel environments with limited training data. 
\end{abstract}

\maketitle

\begin{figure*}[hbt!]
     \centering
         \includegraphics[width=1.0\textwidth]{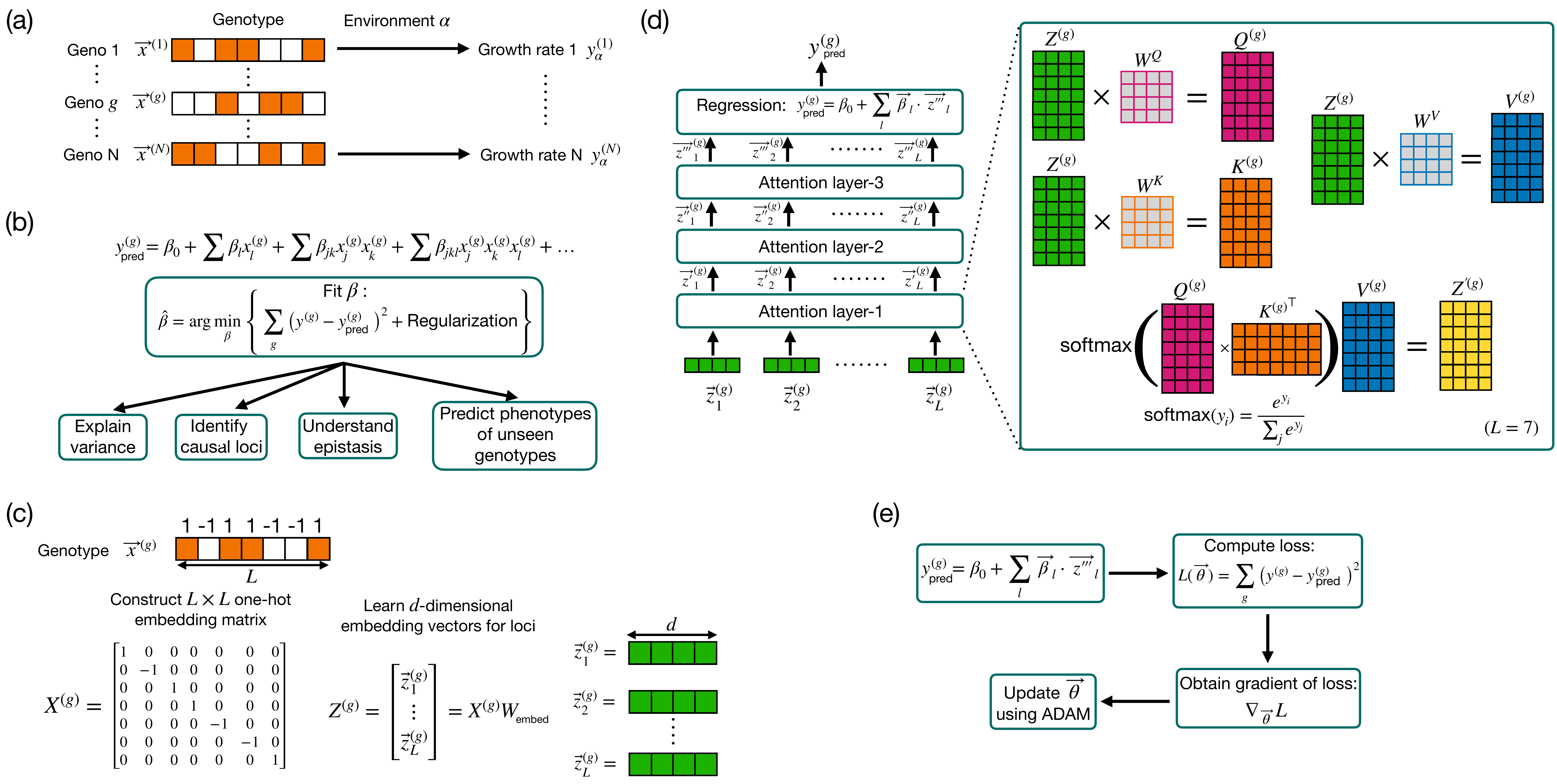}         
    \caption{
        \textbf{Schematic illustrating standard regression-based and attention-based methods for genotype-phenotype mapping.} (a) Genotype sequences represented as vectors $\mathbf{x}^{(g)}$ for the $g$-th individual, with phenotype $y^{(g)}_\alpha$ in environment $\alpha$. (b) Classical series expansion model, combining linear and higher-order epistasis, fitted by minimizing the loss function with regularization. (c) Genotype vectors are converted to one-hot embeddings $X^{(g)}$ and transformed into $d$-dimensional embeddings $Z^{(g)}$. (d) Flowchart illustrating attention-based architecture for a case of $L=7$ loci and $d = 4$. The embeddings pass through multiple attention layers, followed by prediction. (e) The optimization process involves computing the loss $L(\mathbf{\theta})$, obtaining the gradient of the loss $\nabla_{\mathbf{\theta}} L$, and updating the parameters $\mathbf{\theta}$.
    }
        \label{fig_main_schematic}
\end{figure*}

Mapping the genetic basis of complex traits is a central goal in quantitative genetics  \cite{wagner2011pleiotropic, paaby2013many, solovieff2013pleiotropy, rockman2008reverse, mackay2001genetic, manolio2009finding}. This challenge is typically tackled by measuring genotypes and phenotypes of a large number of diverse individuals, and then identifying statistical associations between genetic variants and corresponding phenotypes. Numerous methods have been developed to infer these genotype-phenotype maps using data from genome-wide association studies (GWAS) in humans or from controlled crosses in model organisms (e.g. quantitative trait locus mapping) \cite{visscher2012five, visscher201710, uffelmann2021genome, ba2022barcoded, petti2023inferring}. These existing methods are typically based on linear regression. Each genotype is represented as a vector of allelic states at a series of genetic loci (see Fig.~\ref{fig_main_schematic}(a) and details in the next section). The phenotype corresponding to each genotype can then be parameterized based on the linear effect of each locus, plus pairwise and higher-order epistatic interactions (see Fig.~\ref{fig_main_schematic}(b)). GWAS or quantitative trait locus (QTL) mapping studies fit the parameters of this model by minimizing prediction error across individuals \cite{myles2008quantitative, powder2019quantitative}. The resulting model can then be used to quantify how much of the observed variability in phenotypic traits can be attributed to genetic factors, to identify which genetic loci are most strongly associated with the phenotypes, and to predict the phenotypes of new individuals from their genotype. 

Because the number of individuals in even the largest studies is much smaller than the number of possible genotypes, some form of regularization is typically applied to avoid overfitting. Traditional analyses begin with an additive model, where the expected phenotype is determined by the sum of independent effects at some set of causal loci. Effects of potential pairwise epistatic interactions between these loci can then sometimes also be mapped. However, existing methods lack power to infer more complex patterns of epistatic interactions across multiple loci. The potential impact of this epistasis remains controversial. Variance partitioning methods \cite{yang2011gcta, pazokitoroudi2020efficient}  often show that epistasis explains only a small fraction of observed variance in phenotypes, but this may be in part because the effects of epistasis are treated as a perturbation to an initial fit based on additive effects \cite{krishna2016limitations}. In addition, because biological systems are composed of interconnected and nonlinear networks \cite{boyle2017expanded}, even sparse epistatic effects can provide key insight into the molecular basis of phenotypes. 

Gene-environment interactions are also challenging to model within the context of traditional methods in quantitative genetics \cite{hunter2005gene, smith2008gene}. These occur when the effect of a genetic variant on a phenotype depends on the environmental context. For instance, individuals with a genetic predisposition to obesity might only exhibit this trait if they are exposed to high-calorie diets \cite{llewellyn2015behavioral}. Incorporating these interactions in GWAS is challenging because it requires comprehensive environmental data alongside genetic data, and the potential environmental variables and their interactions with genetic factors can be vast and complex. Typically, a linear mixed-model approach is used to study gene-environment interactions by including interaction terms in analyses, though the true interactions may not always be linear \cite{korte2012mixed, moore2019linear}.

Despite dramatic increases in the scale of GWAS and QTL mapping studies, traditional methods often fail to capture intricate patterns and interactions, prompting the exploration of machine learning (ML) approaches \cite{nicholls2020reaching}. These include convolutional neural networks \cite{zeng2021g2pdeep, zhou2019whole}, graph neural networks \cite{li2024prs}, evolution-informed pipelines \cite{cheng2021evolutionarily}, and more specialized interpretable deep learning architectures \cite{van2021gennet}. Other approaches seek to utilize deep learning-based computer vision models to predict phenotypes using MRI images and facial recognition technology \cite{pirruccello2022deep, dingemans2023phenoscore}

Recent developments in attention-based models present a promising alternative machine learning paradigm for learning genotype-phenotype maps \cite{vaswani2017attention, lin2022survey}. Originally designed for natural language processing, attention-based models excel at capturing context-dependent interactions, a capability particularly relevant for genetic data, where the effect of a locus can be highly dependent on broader genetic and environmental contexts. Attention-based models have been successfully applied beyond natural language processing, notably to predict protein structure and function with remarkable accuracy \cite{jumper2021highly, rives2021biological}, and to predict binding of T cell receptors to epitopes \cite{meynard2024tulip}.

In this study, we apply attention-based models to predict phenotype from genotype. These attention-based models offer several potential advantages over traditional methods. They are highly expressive because they use learned vectors that capture the entire context relevant to the effects of a given locus on phenotype, rather than relying on context-independent scalar effects. This capability enables them to capture subtle genetic interactions. Another key strength is the ability to incorporate environmental tokens naturally within the structure of the model, making it possible to account for potential gene-environment interactions and hence adapt predictions based on the environment. Traditional methods often lack this flexibility. Additionally, attention-based models can learn environment-dependent vectors, and hence interpolate to make predictions for environments not present in the training data. 

The architecture of the attention-based models begins by converting genotypes into vectors, which are then processed through attention layers to capture epistatic and environmental interactions. This approach enables the model to capture effects of complex epistasis, adapt to environmental contexts, and predict unseen genotypes by learning a generalizable mapping from genotype to phenotype. As a proof of principle, we first apply these methods to analyze simulated data in models with varying degrees of epistatic complexity, and then apply our approach to data from a recent QTL mapping study in budding yeast. 

In the next section, we provide a detailed explanation of the attention-based architecture used in this study, highlighting its components and the underlying mechanisms that make it well-suited for G-P mapping. In subsequent sections, we then apply this architecture to analyze simulated data in models of varying epistatic complexity and experimental data from a recent QTL mapping study in budding yeast that quantitatively measured the growth rate of 100,000 individuals across 18 different environments \cite{ba2022barcoded}. This unique study is ideally suited for exploring the efficacy of attention-based architectures which are thought to function best with large amounts of data. Finally, we present a multi-environment generalization of our attention-based approach, and analyze its ability to ``transfer knowledge'' to new environments. 

\section{Attention-based architecture for G-P mapping}
Our ultimate goal is to develop attention-based models that can predict phenotypes based on genotype data, while also accounting for epistasis and gene-environment interactions. Before diving into the details of the attention-based architecture, it is essential to understand the standard regression model commonly used in the field. This model serves as a foundation for understanding G-P mapping.

\subsection{Overview of standard model}
To make the discussion concrete, we focus on the yeast QTL experiments in \cite{ba2022barcoded} where the authors constructed a panel of 100,000 F1 haploid offspring (segregants) from a cross between the laboratory strain BY and the vineyard strain RM. Each segregant was sequenced to determine its genotype, and the relative growth rate of all segregants was measured in each of 18 laboratory media conditions. For our purposes here, this growth rate data represents 18 different phenotypes. 

In this dataset, the two parental strains differ at approximately $L=41,594$ loci (corresponding primarily to single-nucleotide polymorphisms) and exhibit variation in many relevant phenotypes. For any given offspring $g$, we denote the two possible genotypes (corresponding to the BY or RM parent respectively) at locus $l$ by $x_l^{(g)}=\pm 1$. In other words, in this biallelic case the genotype of the $g$-th individual can be represented by an $L$-dimensional vector $\mathbf{x}^{(g)}$, where the $l$-th element in this vector is either $+1$ or $-1$, depending on the allele present at the $l$-th locus. Each of these genotypes corresponds to a set of 18 measured phenotypes, $y^{(g)}_\alpha$, the growth rate in media condition $\alpha$. 

We then model each predicted phenotype for the $g$-th individual, $y_{\text{pred}}^{(g)}$, as:
\begin{eqnarray}
y_{\text{pred}}^{(g)} & = & \beta_0 + \sum_{j} \beta_j x_j^{(g)} + \sum_{j<k} \beta_{jk} x_j^{(g)} x_k^{(g)} \nonumber \\ & & + \sum_{j<k<l} \beta_{jkl} x_j^{(g)} x_k^{(g)} x_{l}^{(g)} + \ldots \label{eq_series_mod} . 
\end{eqnarray}
Here, the $\beta$ coefficients capture the effect sizes of these loci and their epistatic interactions. Each of the phenotypes is modeled independently. The goal of the model is to estimate the coefficients $\beta$ for each phenotype by minimizing the sum of squared differences between the measurement $y^{(g)}$ and the prediction $y_{\text{pred}}^{(g)}$. This is mathematically expressed as:
\begin{equation}
\hat{\beta} = \arg \min_{\beta} \left\{ \sum_{g} \left( y^{(g)} - y_{\text{pred}}^{(g)} \right)^2 + \text{Regularization} \right\} ,
\end{equation}
where the regularization term helps prevent overfitting by penalizing large coefficient values, ensuring that the model remains generalizable. 

The purely linear version of this model (with only $\beta_0$ and $\beta_j$ terms nonzero) assumes that the effects of different loci are additive. More generally, the model can be used to fit both linear and epistatic terms, but data limitations often make it difficult to infer even pairwise epistatic terms. Higher-order terms are in practice almost always impossible to estimate. These limitations lead to incomplete capture of the complexity of the genetic architecture. While the model uses regularization to prevent overfitting, this does not circumvent the fundamental issue of using limited data to explore a high-dimensional genomic space. In addition, standard models struggle to incorporate the impact of environmental factors (e.g. in the case of our yeast data, relationships between the growth rates in the different media conditions), which can significantly influence phenotypic traits. 

\subsection{Attention-based architecture}
To address the limitations of traditional models, we propose an attention-based architecture. We first describe a single-environment version of this architecture that learns the G-P map for each environment separately. Later, we will discuss the multi-environment architecture, which can learn the G-P map for all environments jointly, capturing gene-environment interactions and cross-environment information.

In the subsequent subsections, we describe the key specific components of the single-environment attention-based model, and explain how each part contributes to its capacity to handle complex epistasis. We outline the process from genotype representation to the final prediction of phenotypes and optimization, emphasizing throughout the distinctive features of attention-based architecture (see Fig.~\ref{fig_main_schematic}(c), (d), and (e)).

\subsection{Genotype representation and embedding} 
The first step in our statistical procedure is the transformation of genotype data into a format that can be processed by the attention-based architecture (see Fig.~\ref{fig_main_schematic}(c)). The $L$-dimensional genotype vector $\mathbf{x}^{(g)}$ is encoded as $L$ one-hot vectors with a single non-zero element, which is $\pm 1$. In this encoding, the $l$-th element of the $l$-th one-hot vector is equal to $x^{(g)}_l$, while the remaining elements are zeros. The one-hot embedding matrix $X^{(g)}$, where each row corresponds to the one-hot vector for each locus, can be written as:
\begin{equation}
X^{(g)} = \begin{bmatrix}
1 & 0 & 0 & \cdots & 0 \\
0 & -1 & 0 & \cdots & 0 \\
\vdots & \vdots & \vdots & \ddots & \vdots \\
0 & 0 & 0 & \cdots & 1 \\
\end{bmatrix} .
\end{equation}

This matrix is then transformed into a continuous, dense representation using a learned $L \times d$-dimensional embedding matrix $W_{\text{embed}}$, resulting in an $L \times d$-dimensional matrix $Z^{(g)}$ for each individual:
\begin{equation}
Z^{(g)} = X^{(g)} W_{\text{embed}} .
\end{equation}
The $l$-th row  in $Z^{(g)}$ corresponds to a $d$-dimensional embedding vector $\mathbf{z}^{(g)}_l$ for the $l$-th locus, capturing its genetic information in a form suitable for subsequent processing by the attention-based model. This transformation is useful because the original genotype matrix $X^{(g)}$ is high-dimensional and sparse, making direct use computationally expensive. The embedding matrix reduces dimensionality and captures essential genetic information, leading to more efficient computations.

The learned embedding vectors $\mathbf{z}^{(g)}$ represent complex relationships between genetic loci that might not be apparent in the raw one-hot encoded representation. Learned dense embeddings are more suitable for neural network-based models, like transformers, providing a richer representation of input data \cite{bengio2013representation}. 

The embedding dimension $d$ controls the trade-off between capturing complex genetic interactions and minimizing computational cost. We select $d$ by evaluating validation performance across different values, $d= 2,12,30,50$. The attention-based architectures showed good performance for all choices of $d$ (see Fig.~\ref{fig_syn_schematic}(c)). However, for simulated data where phenotypes are explicitly generated with high-order epistasis and a structured G-P relationship, we found $d=30$ was optimal especially for highly epistatic models. However, for experimental data, where the true G-P map is unknown, our simulations suggested $d=12$  was sufficiently large to capture complex phenotypes while minimizing computational costs. 

\subsection{Attention mechanism}
The core of the attention-based architecture is the attention mechanism, which enables the model to weigh the importance of different loci. In the attention layer, the embeddings are transformed into three sets of vectors: queries $Q^{(g)}$, keys $K^{(g)}$, and values $V^{(g)}$ (see Fig.~\ref{fig_main_schematic}(d)). These transformations are achieved through learned weight matrices $W^Q$, $W^K$, and $W^V$:
\begin{equation}
Q^{(g)} = Z^{(g)} W^Q, \quad K^{(g)} = Z^{(g)} W^K, \quad V^{(g)} = Z^{(g)} W^V .
\end{equation}
The necessity of these three different vectors arises from the design of the attention mechanism, which aims to flexibly capture complex interactions between loci while maintaining computational tractability. The queries $Q^{(g)}$ represent the perspective of the current locus, seeking information about other loci.  The keys $K^{(g)}$ represent the potential loci that the current locus may attend to, effectively serving as points of reference. The values $V^{(g)}$ contain the actual information that should be integrated, based on the relevance determined by queries and keys.

The attention mechanism calculates the relevance between different loci by computing the attention scores, which are determined as the dot product of queries and keys. The output of an attention layer is given by:
\begin{equation}
Z^{'(g)}= \text{softmax}\left(Q^{(g)} {K^{(g)}}^\top\right) V^{(g)} ,
\end{equation}
where the softmax function is defined as:
\begin{equation}
\text{softmax}(y_i) = \frac{\exp(y_i)}{\sum_{j} \exp(y_j)}.
\end{equation}
The softmax function normalizes the attention scores into a probability distribution, emphasizing the most relevant loci while suppressing less relevant ones. The output of the attention layer, therefore, is a weighted sum of the value vectors $V^{(g)}$, where the weights are the normalized attention scores. 

This process relates closely to the concept of epistasis, where the effect of one gene is influenced by others. In the attention mechanism, the query $ Q^{(g)}$ from a given locus interacts with the keys $ K^{(g)} $ of other loci to determine relevance. The softmax function then assigns higher weights to locus pairs with stronger interactions. This adjustment allows the model to capture the epistasis by emphasizing loci combinations that are contextually significant, thus highlighting how different genes influence each other within the genetic network. This capability to focus on important genetic relationships makes the attention-based model particularly effective for modeling epistasis.

Stacking multiple attention layers improves the model's ability to capture higher-order epistatic interactions by iteratively refining how loci influence each other. The first layer learns direct pairwise interactions, while deeper layers integrate signals across three or more loci, capturing complex, non-linear dependencies. Such stacking of attention layers is commonly used in large language models, including those used to model proteins.  We use three layers because they collectively capture both pairwise and higher-order interactions, and empirical tests showed that adding more layers did not improve performance.

\subsection{Prediction}
After the embeddings have been processed through the attention layers, the final representation ${Z'''}^{(g)}$ (obtained from the third attention layer) is used for prediction (see Fig.~\ref{fig_main_schematic}(d)). The predicted phenotype $y^{(g)}_{\text{pred}}$ for each individual $g$ is given by taking these vectors and processing them through a ``regression layer'':
\begin{equation}
y^{(g)}_{\text{pred}} =\beta_0+\sum_l \bm{\beta}_l\cdot \mathbf{z'''}^{(g)}_l . 
\end{equation}
Here, $\bm{\beta}_l$ represents the weights associated with each embedding, and $\beta_0$ is the bias term.  This regression layer is necessitated by the fact that the output of our model is a continuous real-valued number: the predicted growth rate of an individual in a given environment. 

\subsection{Optimization}

The objective is to minimize the prediction error, which is typically quantified by a loss function. A loss function measures the difference between the predicted and actual values of the quantities of interest. We employ mean squared error as our loss function $L(\bm{\theta})$, where $\bm{\theta}$ represents all the parameters involved in the architecture. It is defined as:
\begin{equation}
L(\bm{\theta}) = \sum_g \left(y^{(g)} - y^{(g)}_{\text{pred}}\right)^2 .
\end{equation}
We minimize this function because, theoretically, at its minimum, $y^{(g)}$ equals $y^{(g)}_{\text{pred}}$. 

The entire model is trained end-to-end using gradient-based optimization methods (see Fig.~\ref{fig_main_schematic}(e)).   The process begins with an initial guess for the model parameters $\bm{\theta}_0$. The input embedding vectors are passed through the architecture, and the gradient of the loss function $\nabla_{\bm{\theta}}L(\bm{\theta})$ is calculated. The gradients indicate the direction and magnitude by which each parameter should be adjusted to reduce the loss. The parameters are then updated by moving in the direction of the gradient using gradient descent or more sophisticated second-order methods such as ADAM \cite{kingma2014adam, mehta2019high}, until a local minimum of the loss function is reached.

In this work, the PyTorch library is utilized for its efficient automatic differentiation capabilities \cite{paszke2019pytorch}. PyTorch computes the gradients of the loss function using backpropagation, which applies the chain rule of calculus to propagate the error gradients backward through the network from the output layer to the input layer \cite{mehta2019high}.

\subsection{Implementation}
A detailed explanation of how the attention-based architecture is implemented using PyTorch is given in the Appendix \ref{sec_app_3}. In addition, all code used throughout this work is available on Github at our Github repository \href{https://github.com/Emergent-Behaviors-in-Biology/GenoPhenoMapAttention}{https://github.com/Emergent-Behaviors-in-Biology/GenoPhenoMapAttention}. 

\begin{figure*}[hbt!]
     \centering
         \includegraphics[width=0.84\textwidth]{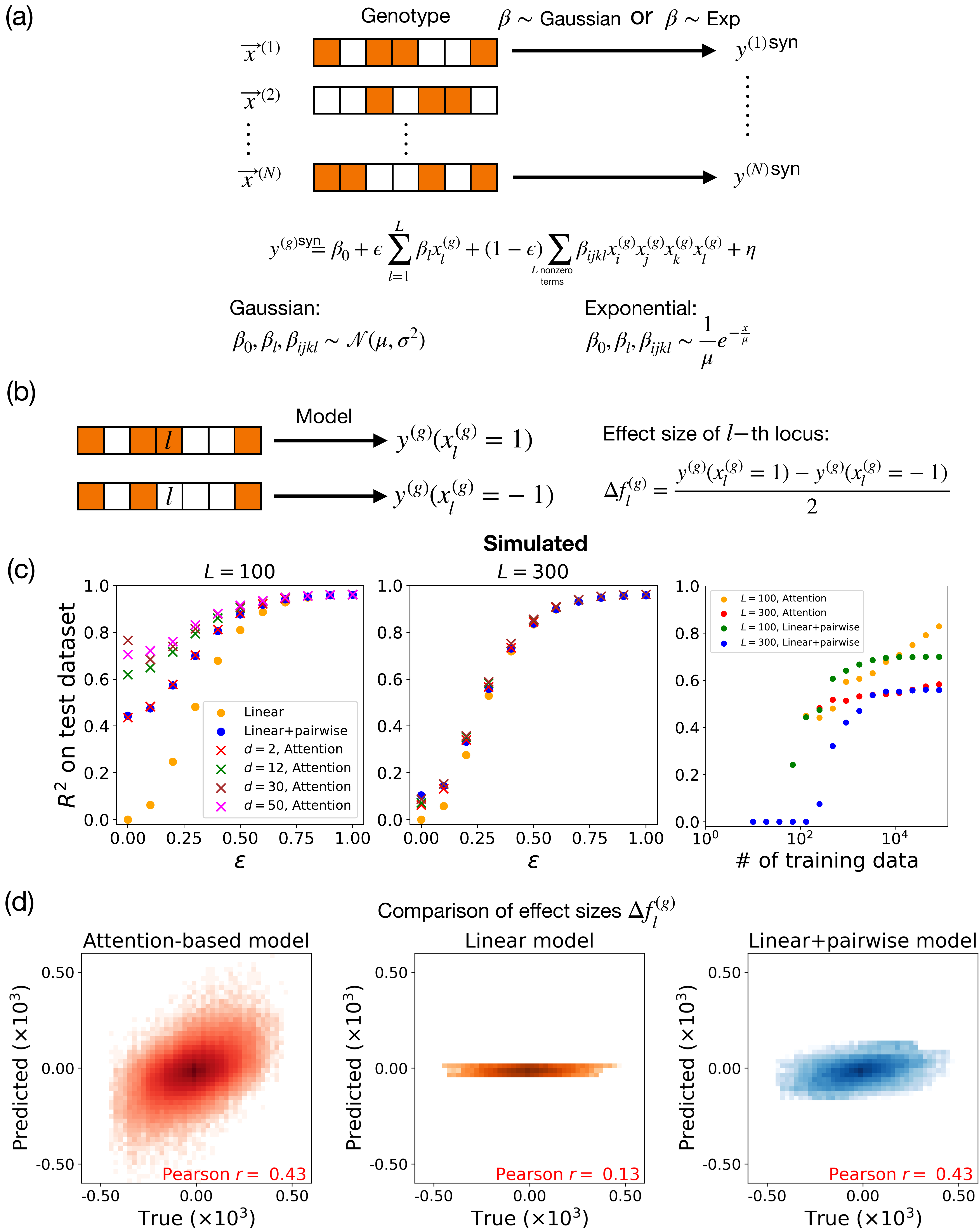}         
   \caption{
        \textbf{Model comparison on synthetic data}. 
        (a) Schematic of pipeline for generating simulated data. We create subsampled genotype vectors $\mathbf{x}^{(g)}$ for individual $g$ from the real genotype data in \cite{ba2022barcoded}, and simulate the phenotype of this individual according to the equation shown, with coefficients drawn from Gaussian (main text) or Exponential (Appendix) distributions. 
        (b) We define the model prediction for the effect size of the $l$-th locus in the $g$-th background genotype as $\Delta f_l^{(g)}$, which is the difference between the predicted phenotype with the $l$-th locus being 1 versus -1 in that genetic background at all other loci. 
        (c) $R^2$ scores for linear, linear+pairwise, and attention-based models across simulated data sets as a function of the simulated strength of epistasis $\epsilon$. Left panel shows the case of $L=100$ causal loci, while middle panel shows the case of $L=300$ causal loci. Right panel shows performance at $\epsilon = 0.3$ for varying training dataset sizes. 
        (d) For $d=30$, $\epsilon = 0.3$, and $L = 100$, the predicted effect sizes for each locus from different models are compared with the true effect sizes. Simulated data are generated using Gaussian-distributed coefficients. }
        \label{fig_syn_schematic}. 
\end{figure*}

\subsection{Parameter complexity comparison}
The attention-based model contains significantly fewer parameters compared to the standard linear plus pairwise model whenever the number of loci is much smaller than the embedding dimension ($L \ll d$). The total number of parameters in the attention-based model is given by $2Ld + 3d^2 N_L + 1$, where $N_L$ is the number of attention layers. In contrast, the standard linear+pairwise model, which accounts for additive effects and pairwise interactions, has $\frac{L(L + 1)}{2} + 1$ parameters.

For example, when $L = 1000$, $d = 12$, and $N_L = 3$, the attention-based model has 25,297 parameters, whereas the linear+pairwise model has 500,501 parameters. This means the linear+pairwise model contains roughly 20 times more parameters than the attention-based model. Despite this lower parameter count, we show below that the attention-based model architecture nevertheless efficiently captures higher-order interactions.

\section{Benchmarking on simulated G-P maps}
To evaluate the ability of our attention-based model to capture complex epistasis, we benchmarked it against traditional methods using synthetic data from a set of simulated genotype-phenotype maps. This approach allows us to systematically test model performance in controlled scenarios with varying level of epistasis. 

As described above, a key challenge in traditional methods is limited capacity to account for complex epistasis. Therefore, to probe whether our attention-based model can more accurately capture these complex interactions, we generated a set of simulated genetic architectures where phenotype is determined by a combination of additive (linear) terms along with fourth-order epistatic interactions (see Eq.~(\ref{eq_simulated}) below). We tune the relative contributions of linear and fourth-order interactions across different simulated architectures, and analyze the performance of traditional versus attention-based models as a function of these relative contributions of simple versus complex epistasis. 

\subsection{Generating simulated data}
A key challenge in inferring genotype-phenotype maps is that the genotype structure of sampled individuals is typically shaped by numerous historical, demographic, and evolutionary processes that are unrelated to the phenotypes we aim to predict. We therefore cannot reliably compare methods by assessing their performance on simulated data with random genotypes. For this reason, to maintain realistic genotype structure, we constructed all our synthetic datasets using the actual observed yeast genomes of the $\sim 100,000$ individuals from the QTL experiments in \cite{ba2022barcoded}. This preserves the linkage structure inherent in the cross used to generate these individuals. 

As discussed in the original paper, these individuals differ at $41,594$ loci across the genome, but because of the tight linkage in these F1 segregants, genotypes at nearby loci are highly correlated. This creates a ``fine-mapping'' problem that is extensively discussed in \cite{ba2022barcoded}. However, our goal here is to focus on inference of complex epistasic and environmental interactions, rather than on fine mapping (where we would not expect attention models to have any natural advantage over other methods). With this in mind, we subsample the loci (effectively combining highly correlated loci) to create a representative set of $L=1,164$ independent loci (see section on applications to data below). Ref. \cite{ba2022barcoded} found that, using traditional methods, several hundred loci could be identified as causal. For this reason, in our synthetic datasets we typically choose $100$ to $300$ random loci we designate to be causal. 
  
To assign a synthetic phenotype to each of these genotypes, we use the approach shown in Fig.~\ref{fig_syn_schematic}(a). As before, the genotype vector $\mathbf{x}^{(g)}$ of the $g$-th individual, with $L$ loci, is an $L$-dimensional vector. In this vector, the $l$-th element is either $+1$ or $-1$, depending on the allele present at the $l$-th locus (in the real genotypes, subsampled as described above). We then assign the phenotype $y^{(g)\text{syn}}$ to each genotype:
\begin{align}
y^{(g)\text{syn}} &= \beta_0 + \epsilon \sum_{l=1}^{L} \beta_l x^{(g)}_l \nonumber\\
&+ (1 - \epsilon) \sum_{\substack{L~\text{nonzero} \\\ \text{terms}}} \beta_{ijkl} x^{(g)}_i x^{(g)}_j x^{(g)}_k x^{(g)}_l + \eta , \label{eq_simulated}
\end{align}
where the sums are over the randomly designated causal loci (see above). By tuning the parameter $\epsilon$, we can generate synthetic data with varying amounts of epistasis, ranging from purely additive ($\epsilon = 1$) to purely higher-order epistatic interactions ($\epsilon = 0$). In the main text, we focus on the case where $\beta$ terms are drawn from a Gaussian distribution $\mathcal{N}(\mu, \sigma^2)$; in Appendix, Fig.~\ref{fig_app_2}, we show analogous results for the case when these coefficients are drawn from an exponential distribution.  The noise term $\eta$ represents random environmental or measurement noise and is modeled as a Gaussian distribution with a mean of zero and a standard deviation equal to 20\% of the standard deviation of the simulated fitness.  

\subsection{Comparison of out-of-sample predictive power}
Fig.~\ref{fig_syn_schematic}(c) shows a comparison of our attention-based model with classical linear regression-based approaches (see Eq.~(\ref{eq_series_mod})) on synthetic data generated as described above. All comparisons are based on $R^2$ in a held-out test dataset (see Appendix \ref{sec_app_3} for details of fitting procedures). Throughout this study, we use out-of-sample predictions to compare model performance. Unlike in-sample predictions, which are made on the data used to train the model, out-of-sample predictions are made on new, unseen data. This allows us to assess how well the model can generalize, a key indicator of its practical utility.   

As can be seen in the left most panel of Fig.~\ref{fig_syn_schematic}(c), for $L=100$ loci, the purely linear model performs well in highly linear settings but degrades as the amount of epistasis increases (i.e. as $\epsilon$ decreases). The linear + pairwise model captures some effects of epistasis, but struggles with highly epistatic genotype-to-phenotype maps (i.e. $\epsilon \ll 1$). The attention-based model consistently outperforms other methods, particularly in epistatic regimes, demonstrating its ability to capture high-order epistasis. We find similar results for synthetic data generated with exponentially distributed $\beta$ (see Appendix, Fig.~\ref{fig_app_2}). However, we note that when we increase the number of causal loci used in the synthetic data to $L=300$, the attention-based and linear regression models have similar predictive power. As discussed below, this likely reflects the fact that as the number of causal loci increases, more data is required to to take advantage of the increased expressivity of attention-based models vis-à-vis regression-based approaches. 

\subsection{Dependence of performance on embedding dimension and amount of data}
The performance of the attention-based model depends on the dimension $d$ of the embedding and the amount of available training data (see Fig.~\ref{fig_syn_schematic}(c)). The optimal dimension is influenced by both the complexity of the data and the number of causal loci $L$. A higher number of loci or more complex data  necessitates a larger embedding dimension to adequately capture the variability and epistasis present in the dataset. However, setting $d$ too high can make it difficult to train the model. Therefore, it is crucial to optimize the embedding dimension using a validation set.  Through experimentation, we found that for simulated data with $L = 100$ or $L = 300$, an embedding dimension of $d = 30$ works well. Therefore, we use $d = 30$ for all simulations with synthetic datasets unless explicitly noted otherwise.

 As shown in the right panel of Fig.~\ref{fig_syn_schematic}(c), the performance of the attention-based model improves consistently as the number of data points increases. In contrast, the performance of the linear+pairwise model tends to plateau when the dataset becomes large, indicating its limited capacity to capture more complex interactions. Notably, although the attention-based model has fewer parameters and is roughly as data-hungry as the linear+pairwise model, it achieves equivalent or even greater predictive performance.

\subsection{Comparison of learned effect sizes} 
Next, we investigated whether the attention-based model and other models are capable of learning the effect sizes of causal loci. To do so, we made use of our synthetic data, where we know the true underlying genotype-phenotype maps. This allows us to compare model predictions to exact effect sizes. In the presence of epistasis, the effect size of a locus depends on the genetic background of an individual. For this reason, we focused on the effect size $\Delta f_l^{(g)}$ of changing locus $l$  in an individual with genetic background $g$ (i.e. the difference in phenotype when $x^{(g)}_l$ is set to +1 and -1 in background $g$):
\begin{equation}
\Delta f_l = \frac{1}{N} \sum_{g=1}^N \frac{y^{(g)}(x^{(g)}_l = 1) - y^{(g)}(x^{(g)}_l = -1)}{2}.
\label{eq_effect_size}
\end{equation}
We focused on the case where the genotype-phenotype maps have both a linear component and moderate amounts of epistasis ($\epsilon=0.3$).
As seen in Fig.~\ref{fig_syn_schematic}(d), the attention-based architectures are particularly effective at capturing context-dependent genetic interactions in the test dataset. The linear+pairwise model also showed good performance at this task, as measured by Pearson correlation. However, as can be seen in the plots, the predicted effect sizes are systematically lower than the true effect sizes. Unsurprisingly, the linear model is unable to capture the background dependent effects, since the predictions do not depend on the genomic backgrounds.  

An alternative quantity of interest is ``average'' effect size of a locus across all genomic backgrounds in the dataset. This quantity is a natural proxy for the importance of a locus when epistatic interactions are limited.  Appendix, Fig.~\ref{fig_app_3} shows a plot of the genome-averaged effect size of loci for attention-based models, compared to a purely linear model and a linear+pairwise model. All models are extremely good at predicting average effect sizes as measured by Pearson correlation (attention: $r=0.82$, linear: $r=0.83$, linear+pairwise: $r=0.87$), with the attention-based models having slightly smaller correlation coefficient. However, as can be seen in the Fig.~\ref{fig_syn_schematic}(d), right panel, unlike attention-based architectures the linear+pairwise model consistently underestimates the effect size of loci with large effects. 

\begin{figure}[t!]
     \centering
         \includegraphics[width=0.5\textwidth]{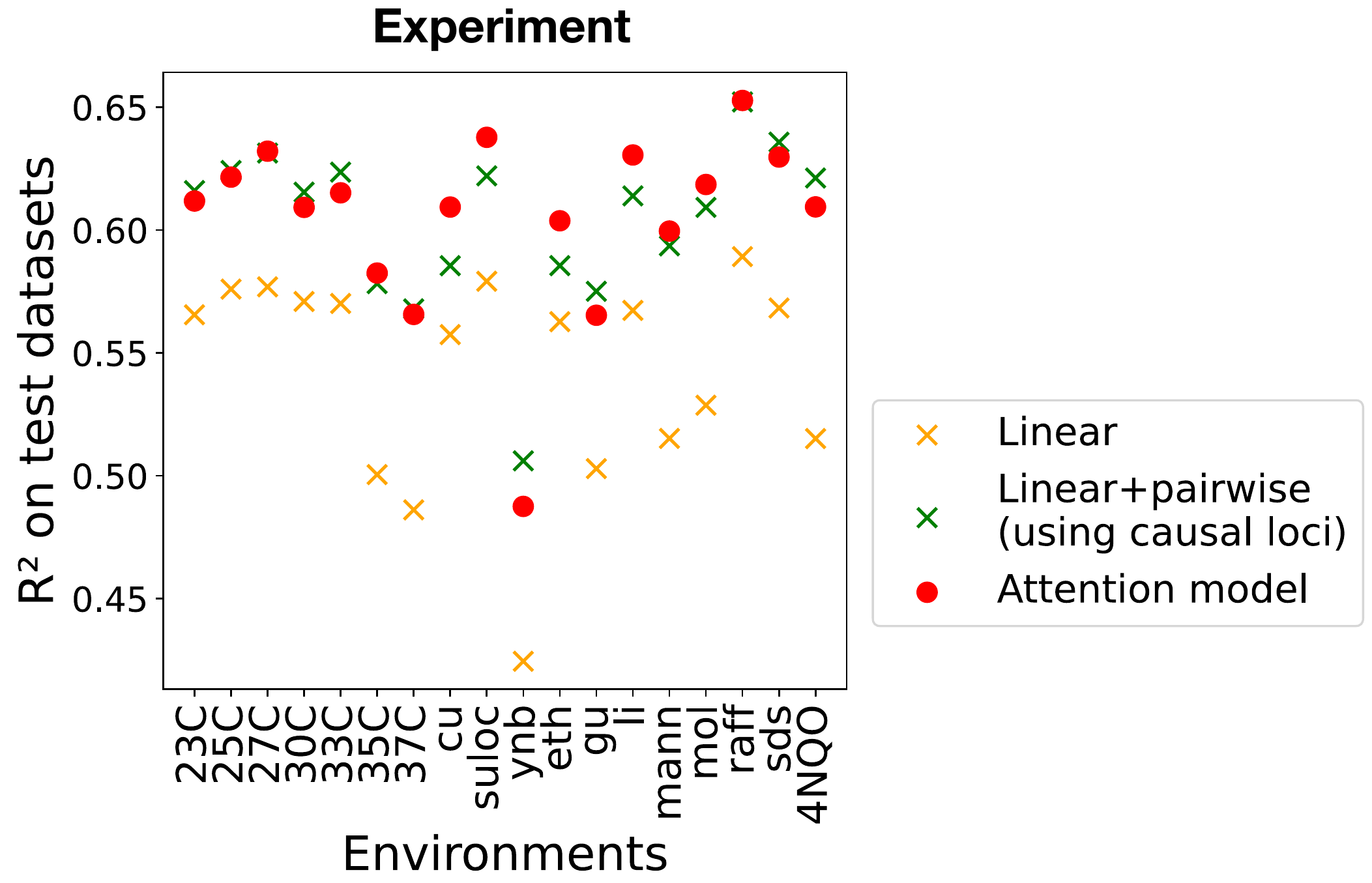}        
    \caption{
        \textbf{Comparison of model performance in yeast QTL mapping data.} We show $R^2$ on test datasets for linear, linear+pairwise, and attention-based model (with $d=12$) across 18 phenotypes (relative growth rates in various environments). For the linear + pairwise mode, the causal loci inferred by \cite{ba2022barcoded} are used.
    }
        \label{fig_single_env}
\end{figure}

\section{Application to inference of empirical G-P maps in budding yeast}

Having benchmarked attention-based models on synthetic data, we next applied these models to the yeast QTL experiments in Ref. \cite{ba2022barcoded}. As described above, this earlier work created a panel of $\sim 100,000$ haploid F1 offspring (segregants) from a cross between a laboratory strain, BY, and a vineyard strain RM. The authors then measured the relative growth rates of these segregants in 18 different laboratory media conditions, including defined minimal media, rich media, several carbon sources, and a range of chemical and temperature stressors (see Ref. \cite{ba2022barcoded} for details on the experimental setup and data acquisition process). 

To analyze this data, we divided the 100,000 segregants into training, validation, and test datasets, with 85\% of the data set aside for training-validation and the remaining 15\% used as the test dataset. The training-validation data were further split into 85\% for training and 15\% for validation. Using the training dataset, we identified $L=1,164$ independent loci, defined as a set of loci such that the correlation between the SNPs present at any pair of loci is less than 94\% (see Appendix \ref{sec_app_1}  for details). 

\subsection{Attention-based model application}
We began by training a separate attention-based model for each of the 18 environments in the dataset (see Fig.~\ref{fig_main_schematic}(d)). Based on hyper-parameter sweeps on the validation set, we chose an embedding dimension of $d=12$ for all models used to analyze experimental data. As expected, the performance of the attention-based model, as characterized by $R^2$ on the test dataset, is much better than that of the linear model (see Fig.~\ref{fig_single_env}). This shows that the attention-based architecture is able to successfully capture the epistatic interactions in the yeast QTL experiments across a wide variety of environments.

Given the increased predictive power of attention-based architectures compared to linear models, we sought to better understand the relationship between the predictions of the linear model and attention-based models.  Fig.~\ref{fig_app_4} in Appendix shows a plot comparing the predicted growth rates of the attention-based models and linear models in all 18 environments for the test dataset. There is strong agreement, with a Pearson correlation coefficients ranging between $r=0.9-0.95$. Fig.~\ref{fig_app_5} in Appendix shows a comparison of the background-averaged effect size of different loci from the attention-based models and linear regression. Once again there is remarkable agreement (Pearson $r=0.83-0.95$). This shows that attention-based architectures can learn all the information contained in linear models while also learning subtle epistatic interactions.

To better understand the kind of epistatic information being captured by attention-based architectures, we also compared our models to a linear plus pairwise model. One major drawback of the linear+pairwise model is that it is computationally infeasible to fit a linear+pairwise model using all $L=1,164$ loci (see discussion of the number of parameters above). For this reason, we restricted our analysis of the linear+pairwise model to the causal loci identified in Ref. \cite{ba2022barcoded}. Focusing on causal loci reduced the number of loci $L$ in our statistical models by an order of magnitude. We found that the linear+pairwise model restricted to causal loci gave similar performance to the attention-based architectures.

Collectively, these results suggest that attention-based architectures can capture subtle epistatic interactions. Moreover, even without any pre-training, the models can automatically focus on the subset of causal loci responsible for these interactions. Our simulations suggest that the performance of the attention-based architectures are limited by the amount of training data. Therefore, it would not be surprising if given more data (or for genotype-phenotype maps with more epistasis), attention-based models outperform linear+pairwise models.

In Appendix, Fig.~\ref{fig_app_6}, we present results on alternative attention based architectures where the genotype loci are represented using one-hot encoding and fed directly into the first attention layer, bypassing any dimensionality reduction. This approach allows the model to process the full-resolution genotype information without compressing it into a lower-dimensional embedding space. In this setup, the dimension of all embeddings is equal to $d=L+1$, where $L$ is the number of loci. We find that for this data regime, these models without dimensionality reduction perform slightly worse than those discussed in the main text.

\begin{figure*}[hbt!]
     \centering
         \includegraphics[width=1.0\textwidth]{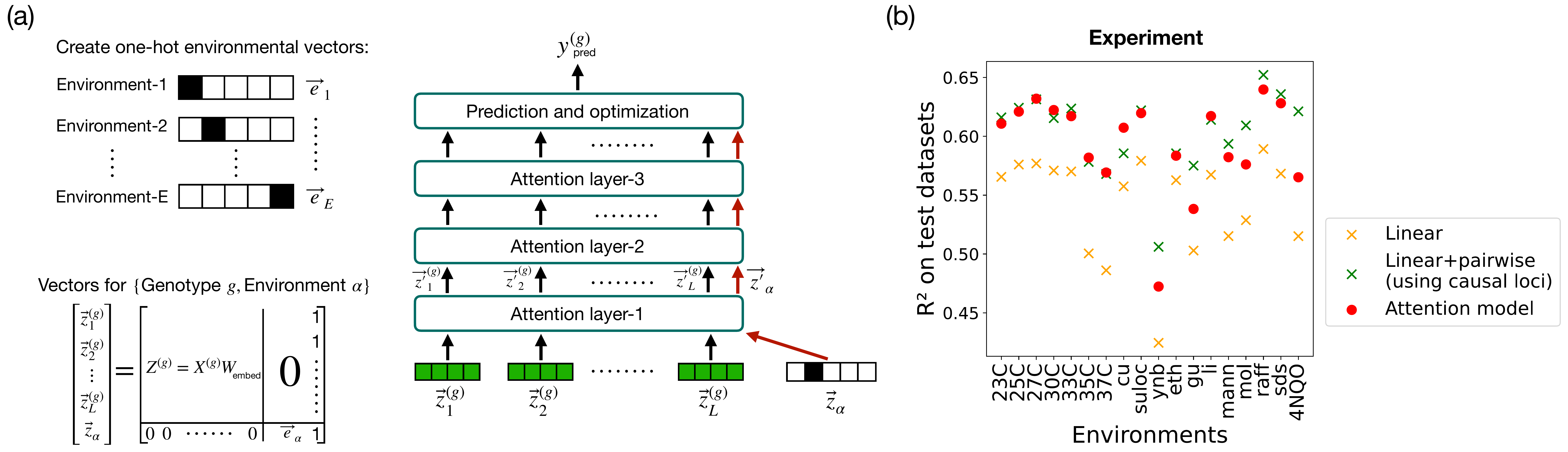}    
    \caption{
        \textbf{Multi-environment attention-based model and performance comparison.}
        (a) Schematic of multi-environment attention-based architecture. One-hot environmental vectors are created for each environment, combined with genotype embeddings $Z^{(g)}$, and processed through multiple attention layers to predict phenotypes $y^{(g)}_{\text{pred}}$.
        (b) $R^2$ performance on test datasets for linear, linear+pairwise, and attention-based model (with $d=12$) across various environments. Note that the linear and linear+pairwise models were trained separately for each environment.
    }
        \label{fig_multi_env}
\end{figure*}

\section{Multi-environment attention-based architecture}
Thus far, we have analyzed each genotype-phenotype map independently. However, phenotypes are often correlated across similar environmental conditions. For example, the growth rate of yeast shows a stereotyped temperature dependence. Standard methods based on linear regression have historically often neglected any information in these correlations, and instead simply infer the genetic basis of each different phenotype independently. More recently, a wide variety of methods have been developed within this general regression-based framework to integrate multiple different phenotypes (e.g. multiple types of xQTL data) into a single joint inference framework.

Attention-based architectures offer a potentially powerful alternative approach to capturing these effects.  In principle, given enough data, a multi-environment attention-based model (i.e. a \emph{single} attention model trained on data from \emph{multiple environments}) should be able share information across environments by incorporating different environmental contexts as additional environmental input vectors (see Fig.~\ref{fig_multi_env}(a)). This opens the possibility of using attention-based mechanisms to transfer information across environments (often called ``transfer learning'' in the machine learning literature \cite{5288526}). In this section, we explore these ideas using both synthetic data and empirical data from the yeast QTL experiments in Ref. \cite{ba2022barcoded}. A more detailed analysis of the performance of multi-environment attention models, including details on construction of the synthetic dataset, model architectures, and training procedure, is provided in Appendix \ref{sec_app_4}.

\subsection{Multi-environment attention-based architecture}
We start by providing a detailed explanation of the multi-environment attention-based architecture used in our studies (see Fig.~\ref{fig_multi_env}(a)).  In this architecture, we introduce additional input vectors to represent the environments. Each environment, indexed by $\alpha$ (where $\alpha \in \{1, 2, \ldots, E\}$), is represented by a one-hot encoded vector, denoted as $\mathbf{e}_\alpha$. This encoding allows the model to distinguish between various environmental conditions while processing genotype data. The locus embeddings  $\mathbf{z}^{(g)}$ have $E$ zeros at the right side, and $\mathbf{e}$ has $L$  zeros at the left side. This ensures that all vectors are of the same length ($d+E$) and have unique representations for each locus and environment. Additionally, a column of ones is added at the right end to enhance the model's capability to capture epistasis. This is motivated by our analytical results (see Appendix \ref{sec_app_6}). The combined genotype and environment vectors are processed through multiple attention layers, capturing interactions between loci within the environmental context. The output from the final attention layer is used to   predict the phenotype $y^{(g)}_{\text{pred}, \alpha}$ and update the model parameters (as shown in Fig.~\ref{fig_main_schematic}(e)).

\subsection{Prediction on yeast data}
The results from applying the model to yeast data are shown in Fig.~\ref{fig_multi_env}(b). This figure compares the performance ($R^2$ on held-out test datasets) of the multi-environment attention-based model with traditional linear and linear+pairwise models across different environments. The multi-environment attention-based model performs better than the single environment linear models and comparably to the single-environment linear+pairwise models. It also has similar predictive power to the more specialized single-environment attention-based models. Our results show that it is possible to train a single model that can capture epistatic interactions across multiple environments. In Appendix, Fig.~\ref{fig_app_6}, we show additional results for alternative different multi-environment attention architectures.

We note that, unlike in protein language models—where attention maps often correlate with structural features \cite{jumper2021highly}—we have found that the attention maps of the learned models are hard to interpret (see Appendix, Fig.~\ref{fig_app_7}). This may reflect the fact that the outputs of our models are continuous rather than discrete, or other complexities of the underlying biology of predicting growth rates from genotype.

\begin{figure*}[hbt!]
     \centering
         \includegraphics[width=1.0\textwidth]{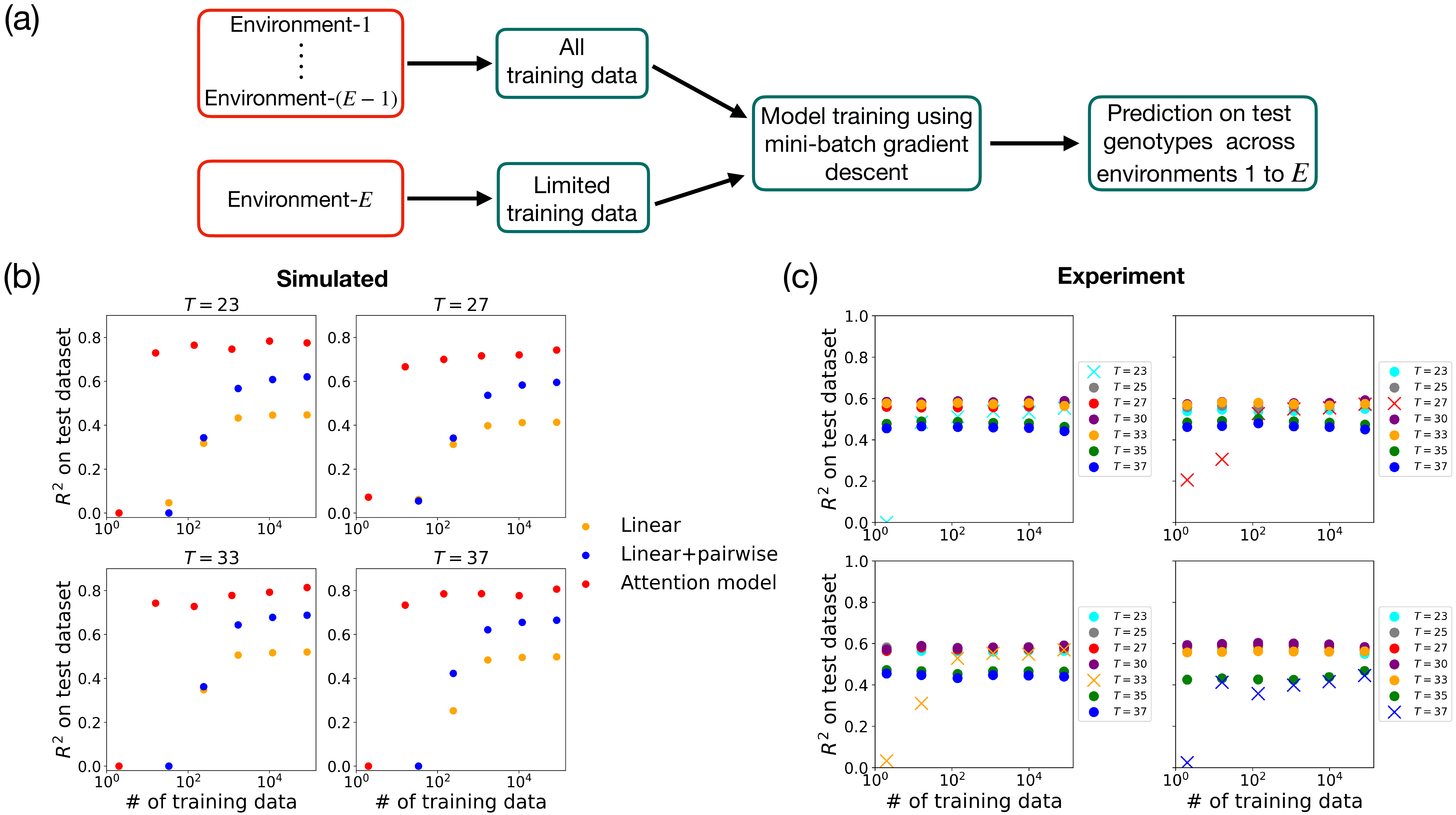}         
\caption{ \textbf{Transfer learning.} (a) Schematic of our approach to transfer learning. Environment-$E$ has fewer training data points than the other environments. Training data from different environments are sampled and used to train the model for phenotype prediction.  (b,c) $R^2$ on the test dataset for (b) simulated data (with $d=30$) and (c) experimental yeast QTL data (with $d=12$). Along the horizontal axes, ``training data'' refers specifically to the number of genotypes from the new temperature $T$ included in the training set; for all other temperatures, the maximum available training set size is used. For the linear and linear+pairwise models in (b), the same number of training data points from temperature $T$ is used. Simulated data are generated using Gaussian-distributed coefficients with $\epsilon=0.3$ and $L=100$.}

        \label{fig_transfer}
\end{figure*}

\subsection{Transfer learning to new environments}
One of the crucial applications of the multi-environment attention-based architecture is its ability to predict phenotypes for environments where data are sparse. This capability to transfer information from data rich environments to new environments with limited data is known as transfer learning (see Fig.~\ref{fig_transfer}(a)). Since attention-based mechanisms are known to excel at  transfer learning, we wanted to ask if our multi-environment architecture could make predictions in a new environment with limited data.

To test this ability, we focused on a subset of seven environments from the yeast QTL experiments in Ref. \cite{ba2022barcoded} where temperature was varied from $23^\circ C$ to $37^\circ C$. We started by constructing a set of synthetic genotype-phenotype maps that mimicked the experimentally measured correlation structure across temperatures (see Appendix \ref{sec_app_2} for details). We then trained
the model on all the training data for six of the temperatures ($\sim 7.2\times 10^4$ individuals), while varying the amount of training data used for the final temperature from a single individual to the full training dataset ($\sim7.2 \times10^4$  individuals). See embedding and training details in Appendix \ref{sec_app_5}. The results are shown in Fig.~\ref{fig_transfer}(b). We found that with as little as $100$ training data points, the multi-environment architecture can successfully transfer information to new temperatures. This shows that our multi-environment attention model is capable of transfer learning on our synthetic dataset.

Having established that transfer learning is possible on synthetic data, we repeated this exercise on data from the QTL experiments in Ref. \cite{ba2022barcoded}. As can be seen in Fig.~\ref{fig_transfer}(c), the multi-environment attention-based model was once again able to learn genotype-to-phenotype maps in a new temperature with as little as $100$ training data points. These results suggests that transfer learning may be a generic feature of attention-based models for genotype-phenotype maps. In contrast, there is no natural way to implement such a transfer learning approach within the context of standard regression-based methods. 

\section{Discussion}
In this paper, we explored the use of attention-based architectures for learning complex genotype-phenotype maps. We demonstrated the potential of these models on both synthetic datasets and experimental yeast QTL data \cite{ba2022barcoded}. Our attention-based models successfully learn complex epistatic interactions. We then explored a multi-environment attention-based model that can predict phenotypes across multiple environments. The multi-environment architecture is specifically designed to leverage shared information across a wide range of environmental conditions, allowing the model to integrate genetic and environmental data into a unified framework. We demonstrated that this multi-environment attention model is capable of ``transfer learning'' genotype-phenotype maps in new environments with limited data. Collectively, our results show that attention-based architectures offer a powerful new class of models for learning genotype-phenotype maps.

The logic underlying our attention-based architectures is fundamentally different from traditional regression-based methods used for learning genotype-phenotype maps. Methods based on linear regression commonly parameterize the genetic architecture of complex traits in terms of an additive model, where the effects of loci are independent. Less often, these models also include pairwise epistatic interactions between loci. Unlike these traditional methods, attention-based models make no assumptions about how loci interact. Instead, they exploit the expressivity of the attention mechanism to directly learn these epistatic interactions from data with minimal assumptions.

There are numerous interesting potential extensions and  future research directions using attention-based models. These include exploring alternative architectures for how genetic and environmental tokens interact and the use of non-linear MLP layers and skip connections. In addition, it will be interesting to apply attention-based architectures to experimental datasets with more demographic structure and genomic diversity. In the yeast QTL dataset analyzed here, all 100,000 individuals are from a single cross between a lab and vineyard yeast strain. This limits both the complexity of the genomic backgrounds and the number and frequency distribution of alleles across loci. We expect attention-based models to perform even better on these richer and more complex datasets. 

Attention-based architectures also open the possibility of directly learning the genomic representations and demographic structure of populations from data at the same time as one learns the genotype-phenotype map. One potential method for doing this is to randomly mask loci during training, analogous to certain large language models such as BERT \cite{devlin2019bert}. Such a procedure would force the attention-architecture to learn correlations between loci present in the training data set without the need for extensive preprocessing.

More generally, our work suggests that as the amount of sequence data increases, it should be possible to use more expressive statistical models to learn genotype-phenotype maps. This opens up the exciting possibility that we may be able to harness advances in machine learning to learn subtle epistatic and gene-environment interactions directly from data. These techniques should also allow us to better understand how environments shape and modulate  genotype-phenotype maps.

\section{Acknowledgements} We would like to thank the Mehta group and Desai lab for helpful discussions. P.M. and K.R. were funded by NIH NIGMS Grant No. R35GM119461 and a Chan-Zuckerburg Institute Investigator grant to P.M. MMD acknowledges support from grant PHY-1914916 from the NSF and grant GM104239 from the NIH. G.R acknowledges support from a joint research agreement between NTT Research Inc. and Princeton University.

\bibliography{gp_transformers}
\clearpage
\widetext

\appendix
\renewcommand{\thefigure}{A\arabic{figure}}
\renewcommand{\theHfigure}{A\arabic{figure}}
\setcounter{figure}{0} 

The code used in this work is available at our GitHub repository 
\href{https://github.com/Emergent-Behaviors-in-Biology/GenoPhenoMapAttention}{https://github.com/Emergent-Behaviors-in-Biology/GenoPhenoMapAttention}. 
We implement our models using the PyTorch framework \cite{paszke2019pytorch}. In the following sections, we detail the procedures and numerical parameters used to generate synthetic data, implement the attention-based models, and perform transfer learning.

\section{Filtering loci for reduced correlation \label{sec_app_1}}

Genetic variants at different loci can be correlated due to linkage disequilibrium, where nearby SNPs tend to be inherited together. To account for this, we first shuffle the genotype data and split it into training ($85\%$) and test ($15\%$) datasets. Next, we compute a correlation matrix $C$, a symmetric matrix where each entry $C_{i,j}$ represents the correlation coefficient between the $i$-th and $j$-th SNPs across individuals in the training dataset. The matrix has dimensions $41,594 \times 41,594$, and its values range from $-1$ (perfect negative correlation) to $1$ (perfect positive correlation). The correlation matrix is computed as:
\begin{equation}
C_{i,j} = \frac{\displaystyle\sum_{g=1}^{N} (x^{(g)}_i - \bar{x}_i)(x^{(g)}_j - \bar{x}_j)}
{\sqrt{\displaystyle\sum_{g=1}^{N} (x^{(g)}_i - \bar{x}_i)^2} \sqrt{\displaystyle\sum_{g=1}^{N} (x^{(g)}_j - \bar{x}_j)^2}} ,
\end{equation}
where $N$ is the number of individuals in the training dataset, $C_{i,j}$ is the Pearson correlation coefficient between the $i$-th and $j$-th SNPs, $x^{(g)}_i$ is the genotype value of the $g$-th individual at the $i$-th SNP, and $\bar{x}_i$ is the mean genotype value at the $i$-th SNP across all individuals in the training dataset:
  \begin{equation}
  \bar{x}_i = \frac{1}{N} \sum_{g=1}^{N} x^{(g)}_i .
  \end{equation}

This correlation matrix allows us to quantify the dependencies between SNPs and subsequently filter loci with an absolute correlation coefficient $|C_{i,j}| > 0.94$, ensuring that only a representative subset remains for further analysis. The procedure follows these steps:

\begin{itemize}
    \item \textbf{Initialize:} 
    \begin{itemize}
        \item Define a set of remaining loci, $R$, containing all loci.
        \item Define an initially empty set, $S$, to store selected independent loci.
    \end{itemize}
    
    \item \textbf{Iterative selection:} 
    \begin{itemize}
        \item Randomly select a locus $i$ from $R$ and add it to $S$.
        \item Identify all loci $j \in R$ that have an absolute correlation $|C_{i,j}| > 0.94$ with the selected locus $i$.
        \item Remove these highly correlated loci from $R$.
                \item Repeat the process (selecting a new locus from $R$, identifying correlated loci, and removing them) until no loci remain in $R$.
    \end{itemize}
    
    \item \textbf{Output:} The final set $S$ contains 1,164 loci.
\end{itemize}
This process filters out highly redundant loci (with pairwise correlations above 0.94), yielding a representative subset for further analysis.

\section{Generation of simulated phenotype data \label{sec_app_2}}
Simulated fitness values are generated using real genotype structures from 100,000 offspring of a budding yeast cross analyzed in Ref. \cite{ba2022barcoded}. The genotype data consists of 41,594 SNPs in total. To reduce redundancy, we select SNPs with a pairwise correlation of 94\% or less, resulting in a filtered set of 1,164 SNPs (see Section \ref{sec_app_1}). From this subset, we pick either $L=100$ or $L=300$ for analysis. The simulated phenotype generation process is detailed below.

\begin{itemize}
    \item \textbf{Temperature range.} Simulated fitness values are generated for the following temperature values:
        \begin{equation}
                   T \in [23, 25, 27, 29, 31, 33, 35, 37]. \label{eq_temps} 
        \end{equation}

    \item \textbf{Linear and epistatic effects.} The fitness values are modeled as a combination of linear effects and fourth-order epistatic interactions:
    \begin{equation}
    y^{(g)\text{syn}} = \beta_0 + \epsilon \sum_{l=1}^{L} \beta_l x^{(g)}_l 
    + (1 - \epsilon) \sum_{\substack{L~\text{nonzero} \\ \text{terms}}} \beta_{ijkl} x^{(g)}_i x^{(g)}_j x^{(g)}_k x^{(g)}_l + \eta, 
    \label{eq_simulated_app}
    \end{equation}
    where $\beta_0$ is the intercept term, $\epsilon$ controls the proportion of linear vs. epistatic contributions, $\beta_l$ represents the effect of the $l$-th locus, $\beta_{ijkl}$ captures the fourth-order epistatic interaction effects, $x^{(g)}_l$ denotes the genotype of individual $g$ at locus $l$, and $\eta$ represents a noise term drawn from a Gaussian distribution. In Eq.~(\ref{eq_simulated_app}), fourth-order interaction terms are generated by selecting $L$ unique sets of four loci at random.

\item \textbf{Temperature dependence.} The genotype effect coefficients are modeled as quadratic functions of temperature:
    \begin{equation}
        \beta_l(T) = a_l (T - T_0)^2 + b_l (T - T_0) + c_l ,
    \end{equation}
    \begin{equation}
        \beta_{ijkl}(T) = a_{ijkl} (T - T_0)^2 + b_{ijkl} (T - T_0) + c_{ijkl} .
    \end{equation}
Here, $T_0 = 30$ represents the reference temperature. The coefficients $a$, $b$, and $c$ are sampled either from a normal distribution with mean $\mu = 0.5$ and standard deviation $\sigma = 0.5$:
    \begin{equation}
        a, b, c \sim \mathcal{N}(\mu = 0.5, \sigma = 0.5) ,
    \end{equation}
    or from an exponential distribution with a mean of $1$:
    \begin{equation}
        a, b, c \sim \text{Exp}(\lambda = 1) .
    \end{equation}

\begin{figure}[hbt!]
     \centering
         \includegraphics[width=0.4\textwidth]{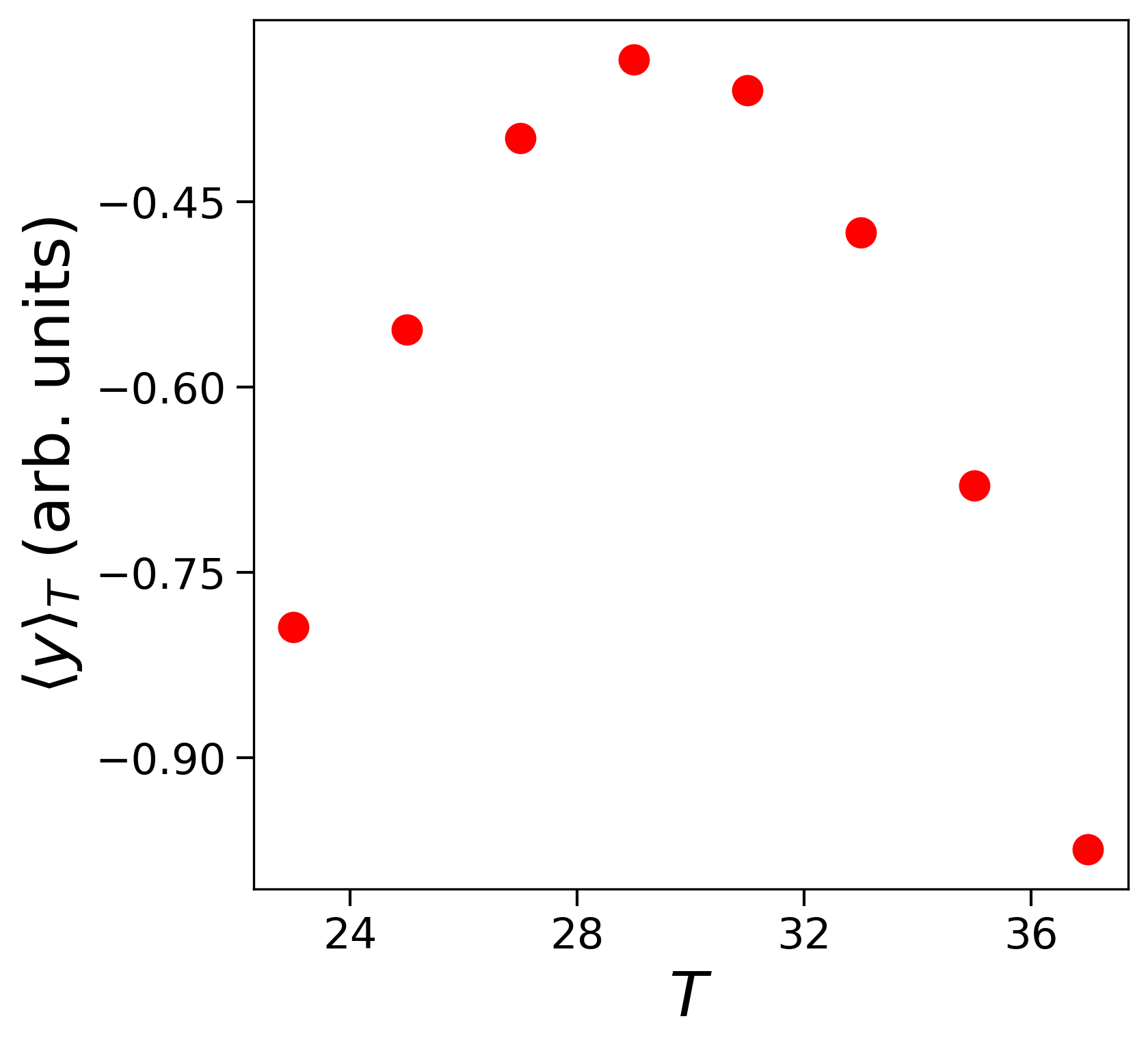}        
    \caption{
       Mean simulated fitness, $\langle y \rangle_T$, as a function of temperature $T$. 
    }
        \label{fig_app_1}
\end{figure}

\item \textbf{Scaling.} Fitness values are then scaled by a factor of $10^{-2}$ to ensure numerical stability and maintain meaningful ranges. The final scaled fitness value is given by:
    \begin{equation}
        y' = -10^{-2}  y ,
    \end{equation}
    where $y$ is the original fitness value before scaling. The negative scaling is applied to ensure that the mean fitness exhibits a peak at $T = T_0$ rather than a minimum, aligning with biological expectations.

\item \textbf{Noise addition.} To simulate measurement variability, random Gaussian noise is added to the generated fitness values. The noise $\eta_{\alpha}$ follows a normal distribution with mean $0$ and a standard deviation set to $20\%$ of the standard deviation of the fitness distribution in its respective environment:
\begin{equation}
    \eta_{\alpha} \sim \mathcal{N} \left( 0, 0.2 \cdot \sigma_{\alpha} \right),
\end{equation}
where $\sigma_{\alpha}$ is the standard deviation of fitness values in environment $\alpha$.
\end{itemize}

Fig.~\ref{fig_app_1} shows the mean simulated fitness values, $\langle y \rangle_T$, as a function of temperature $T$. The fitness exhibits a clear peak around $T = 30$, reflecting an optimal growth condition. This peaked behavior is biologically relevant, as many organisms exhibit temperature-dependent fitness landscapes, with growth rates declining at temperatures too far from their optimal range.  

\section{Implementation of the single-environment attention model \label{sec_app_3}}  

\subsection{Model architecture}

The implemented model consists of three sequential attention layers, each designed to learn genotype-phenotype mappings by attending to specific loci interactions \cite{vaswani2017attention}. To efficiently process high-dimensional genotype data, the model first applies a dimensionality reduction step before passing the embedded input through the initial attention layer. The key components of the architecture are as follows:

\begin{itemize}
    
\item \textbf{Input dimensionality and projection:}
    \begin{itemize}
   \item The genotype data is initially represented in the range $[0,1]$. We first transform it to the range $[-1,1]$ using the transformation:

\begin{equation}
    x^{(g)}_i \leftarrow 2 x^{(g)}_i - 1,
\end{equation}

where $x^{(g)}_i$ is the genotype value for the $g$-th individual at the $i$-th SNP.

\item Each genotype of length $L$ is converted into a one-hot embedding matrix $X^{(g)}$ of shape $L \times L $. The transformation follows:

\begin{equation}
    X^{(g)}_{i,j} =
    \begin{cases} 
        x^{(g)}_i, & \text{if } i = j \\
        0, & \text{otherwise}
    \end{cases},
\end{equation}

where $X^{(g)}$ is a diagonal matrix with genotype values $x^{(g)}_i$ on the diagonal. 

\textbf{Implementation in PyTorch:}

 This transformation can be efficiently implemented in PyTorch as follows:

\begin{verbatim}
# Create a zero matrix of shape (batch_size,L,L):
one_hot_mini_batch_input = torch.zeros((batch_size, L,  L))

# Assign genotype values to the diagonal:
indices = torch.arange(L)
one_hot_mini_batch_input[:, indices, indices] = mini_batch_input.squeeze()
\end{verbatim}
Here, \texttt{mini\_batch\_input.squeeze()} is a tensor of shape $\texttt{batch\_size} \times L$, where each row corresponds to an individual, and each element in a row represents the genotype value at a given locus for that individual.

\item The model processes \texttt{batch\_size} number of genotype matrices $X^{(g)}$, forming a three-dimensional tensor of shape $\texttt{batch\_size} \times L \times L$. 
\item Since the loci embedding vectors can be high-dimensional (e.g., $L=1,164$ in the case of the BBQ budding yeast dataset), our architecture first applies a learned linear projection to map the input into a lower-dimensional latent space of size $d$. This transformation is performed as:
\begin{equation}
    Z^{(g)} = X^{(g)} W_{\text{embed}}, \label{eq_Zg}
\end{equation}
where  $W_{\text{embed}} \in \mathbb{R}^{L \times d}$ is a learnable projection matrix. For experimental data, we set $d=12$, while for synthetic data, we use $d=30$.
\item Dimensionality reduction is not strictly necessary if computational resources and runtime are not a constraint. In such cases, we have also experimented with using the full-dimensional representation, where:

\begin{equation}
    Z^{(g)} = X^{(g)}.
\end{equation}
However, unless specified otherwise, we will focus only on the case where dimensionality reduction is applied.
\item Motivated by the series expansion results (see Section \ref{sec_app_6}), we append a constant value of $1$ as an additional column to each transformed loci representation, effectively extending the embedding dimension from $d$ to $d+1$. The transformation is given by:

\begin{equation}
    \bar{Z}^{(g)} = \begin{bmatrix} Z^{(g)} & \mathbf{1} \end{bmatrix}, \label{eq_Z_with1}
\end{equation}

where $Z^{(g)} \in \mathbb{R}^{L \times d}$ is the original embedding matrix for the $g$-th individual, and the appended column $\mathbf{1} \in \mathbb{R}^{L \times 1}$ consists of ones, resulting in an extended embedding matrix $\bar{Z}^{(g)} \in \mathbb{R}^{L \times (d+1)}$. This can be efficiently performed using \texttt{torch.cat} to concatenate a column of ones created with \texttt{torch.ones}.
\end{itemize}

\item \textbf{Query-Key-Value mechanism and attention computation:}
\begin{itemize}
    \item Each attention layer learns three parameter matrices—query, key, and value—for the following transformations:
    \begin{align}
        Q^{(g)} &= \bar{Z}^{(g)} W^Q, \quad Q^{(g)} \in \mathbb{R}^{L \times (d+1)}, \\
        K^{(g)} &= \bar{Z}^{(g)} W^K, \quad K^{(g)} \in \mathbb{R}^{L\times (d+1)}, \\
        V^{(g)} &= \bar{Z}^{(g)} W^V, \quad V^{(g)} \in \mathbb{R}^{L\times (d+1)},
    \end{align}
    where $W^Q$, $W^K$, and $W^V$ are learnable projection matrices.

    \item The attention scores are computed using a scaled dot-product similarity between query and key matrices:
    \begin{equation}
        A^{(g)} = \text{softmax} \left( Q^{(g)} {K^{(g)}}^T \right). \label{eq_atten_score}
    \end{equation}

    \item The final attended output is obtained by weighting the values using the computed attention scores:
    \begin{equation}
        Z'^{(g)} = A^{(g)} V^{(g)}.
    \end{equation}
    The attention scores are computed using \texttt{torch.matmul} for matrix multiplication, followed by \texttt{torch.softmax} for normalization.

   \item This process is repeated across three stacked attention layers, allowing the model to iteratively refine its understanding of genotype-phenotype relationships. The output of each layer serves as the input to the next layer. Specifically, the output of the first attention layer is denoted as $Z'^{(g)}$ and is used as input for the second layer, producing $Z''^{(g)}$, which in turn serves as input for the third layer, resulting in the final output $Z'''^{(g)}$.
\end{itemize}

\item \textbf{Attention layers and final output computation:}
\begin{itemize}
    \item The model consists of three sequential attention layers, each with its own set of learnable weight matrices.
    \item After the embeddings have been processed through the attention layers, the final representation ${Z'''}^{(g)}$ (obtained from the third attention layer) is used for prediction. The predicted phenotype $y^{(g)}_{\text{pred}}$ for each individual $g$ is given by:

\begin{equation}
y^{(g)}_{\text{pred}} =\beta_0+\sum_l \boldsymbol{\beta}_l\cdot \mathbf{z'''}^{(g)}_l .
\end{equation}
Here, $\bm{\beta}_l$ represents the weights associated with each embedding, and $\beta_0$ is the bias term.
\end{itemize}

\item \textbf{Parameter initialization:}
\begin{itemize}
   \item All learnable weight matrices are initialized using values drawn from a Gaussian distribution to ensure stable training. Specifically, each $W$ is initialized as:

\begin{equation}
    W \sim \mathcal{N}(0, \sigma^2),
\end{equation}

where $\sigma$ varies depending on the specific architecture. In our work, $\sigma$ ranges from $0.01$ to $0.04$, chosen based on empirical performance on the validation set to ensure stability and optimal convergence.

The key learnable weight matrices in the architecture include:
\begin{itemize}
    \item \textbf{Embedding projection:} $W_{\text{embed}}$.
    \item \textbf{Query-Key-Value transformations:} $W^Q$, $W^K$, $W^V$ — different for each layer.
    \item \textbf{Final linear transformation after attention:} $\beta_0$, $\bm{\beta}_l$.
\end{itemize}

\end{itemize}

\end{itemize}

\subsection{Training and optimization}

\begin{itemize}

    \item \textbf{Data splitting:}
    \begin{itemize}
        \item A total of $85\%$ of the samples are allocated to training and validation, while the remaining $15\%$ is reserved as the test dataset.
        \item Within the training-validation set, $85\%$ of the data is used for training, and $15\%$ for validation.
    \end{itemize}

    \item \textbf{Data normalization:}
    \begin{itemize}
        \item Fitness values $y^{(g)}$ are standardized based on the mean and standard deviation of the training set to ensure numerical stability:
        \begin{align}
            \texttt{train\_mean} &= \frac{1}{N} \sum_{g=1}^{N} y_{\text{train}}^{(g)}, \\
            \texttt{train\_std} &= \sqrt{\frac{1}{N} \sum_{g=1}^{N} \left(y_{\text{train}}^{(g)} - \texttt{train\_mean} \right)^2}.
        \end{align}
        \item The training, validation, and test fitness values $y^{(g)}$  are normalized as:
        \begin{equation}
            \hat{y}^{(g)} = \frac{y^{(g)} - \texttt{train\_mean}}{\texttt{train\_std}}, \label{eq_appen_norm}
        \end{equation}
        where $\hat{y}^{(g)}$ denotes the normalized fitness values.
    \end{itemize}

    \item \textbf{Training procedure:}
    \begin{itemize}
        \item The model is trained using mini-batch gradient descent with a batch size of $64$. During training, the first $64$ samples are processed together in one batch, followed by the next $64$, and so on, until the entire training dataset is covered in multiple iterations. The choice of $64$ balances computational efficiency and gradient stability—small enough to fit within memory constraints while large enough to ensure smooth gradient updates and stable convergence.

    \item Training is performed for more than $200$ epochs, meaning the model iterated over the entire dataset at least $200$ times. This ensures that the model has sufficient opportunities to learn complex genotype-phenotype relationships and reach a stable state where further training does not significantly improve performance.

\item The mean squared error (MSE) is used as the loss function, defined as:
    \begin{equation}
        \text{MSE} = \frac{1}{N} \sum_{i=1}^{N} (\hat{y}^{(g)} - \hat{y}^{(g)}_{\text{pred}})^2,
    \end{equation}
    where $\hat{y}^{(g)}$ represents the true normalized phenotype, $\hat{y}^{(g)}_{\text{pred}}$ is the predicted normalized  phenotype, and $N$ is the number of training samples in a mini-batch. 

        \textbf{PyTorch syntax:} \texttt{loss = torch.nn.MSELoss()}

\item The Adam optimizer is used for training, with the learning rate depending on the genotype representation method. When dimensionality reduction is applied to loci embeddings, a learning rate of $0.001$ is used, as it provides a good balance between convergence speed and stability, avoiding large oscillations in weight updates. However, when the original one-hot embedding of dimension $L$ is used, a smaller learning rate of $0.0001$ is chosen to ensure stable training and prevent large updates due to the high-dimensional input space.

    \textbf{PyTorch syntax:}
    \begin{itemize}
        \item For dimensionality-reduced embeddings:
        \texttt{optimizer = torch.optim.Adam(attention\_layer.parameters(), lr=0.001)}
        \item For original one-hot embeddings of dimension $L$:
        
        \texttt{optimizer = torch.optim.Adam(attention\_layer.parameters(), lr=0.0001)}
    \end{itemize}

    The learning rate selection is based on empirical performance on the validation set, ensuring optimal convergence while preventing instability or slow training.

\item Before each epoch, the dataset is shuffled to ensure that each mini-batch contains diverse samples, reducing bias and preventing the model from learning spurious correlations that might arise due to the order of data. Gradient updates are then applied iteratively on these mini-batches using backpropagation. The training process follows these key steps:

\begin{itemize}
    \item Zero the accumulated gradients from the previous iteration using:
    \texttt{optimizer.zero\_grad()} 
    
    \item Compute the gradient of the loss function with respect to model parameters using backpropagation:
    
    \texttt{train\_loss.backward()}
    
    \item Update the model parameters using the optimizer:
    \texttt{optimizer.step()}
\end{itemize}

These steps allow the model to iteratively minimize the loss function and improve generalization.

\end{itemize}

\item \textbf{Validation and model selection:}
\begin{itemize}
\item The validation dataset is used to monitor model performance at the end of each epoch, helping to assess how well the model generalizes to unseen data.
\item At the end of training, the epoch that achieved the highest $R^2$ on the validation set is identified, and the model parameters from that epoch are saved as the best-performing version.
\item This saved version is then restored and used for the final evaluation on the test dataset.
\end{itemize}

\item \textbf{Test performance evaluation:}
\begin{itemize}
\item After training, the model is evaluated on the test dataset to assess its generalization ability.  
\end{itemize}

\item \textbf{Memory optimization:}
\begin{itemize}
    \item To prevent memory overload, phenotype predictions are processed in smaller batches, specifically in chunks of $100$ segregants. This approach ensures efficient memory usage while maintaining computational speed.
    \item GPU memory usage is actively managed by clearing unused tensors with \texttt{torch.cuda.empty\_cache()} between computations. This helps free up allocated memory, reducing the risk of out-of-memory errors and ensuring smoother model execution.
\end{itemize}

\end{itemize}

\section{Multi-environment attention-based model \label{sec_app_4}}

This architecture differs from the single-environment attention model in two key aspects: the embedding representation and the mini-batch selection strategy. Here, we focus on these differences.

\subsection{Encoding genotypes and environments}

In the multi-environment model, each environmental condition is represented by a one-hot encoded vector of dimension equal to the total number of environments, $E$. The entry corresponding to the environment’s index is set to $1$, while all other entries are $0$. These environmental embeddings are processed alongside the genotype embeddings within the attention-based architecture, allowing the model to learn environment-specific phenotypic effects without explicit concatenation.

To ensure compatibility within the attention mechanism:
\begin{itemize}
    \item The $d$-dimensional loci embeddings $Z^{(g)}$ (see Eq.~(\ref{eq_Zg})) are padded with $E$ zeros on the right:
    \begin{equation}
        \bar{Z}^{(g)} = \begin{bmatrix} Z^{(g)} & \mathbf{0}_{L \times E} \end{bmatrix}, \quad \bar{Z}^{(g)} \in \mathbb{R}^{L \times (d+E)}.
    \end{equation}
    
    If no dimensionality reduction is applied (i.e., the full loci embedding size $L$ is used), the loci embeddings are padded as:
    \begin{equation}
        \bar{Z}^{(g)} = \begin{bmatrix} Z^{(g)} & \mathbf{0}_{L \times E} \end{bmatrix}, \quad \bar{Z}^{(g)} \in \mathbb{R}^{L \times (L+E)}.
    \end{equation}
    
    \item The $E$-dimensional one-hot encoded environment embeddings $\mathbf{e}_\alpha$ are padded with $d$ zeros on the left:
    \begin{equation}
        \mathbf{\bar{e}}_\alpha= \begin{bmatrix} \mathbf{0}_{1 \times d} & \mathbf{e}_\alpha \end{bmatrix}, \quad \mathbf{\bar{e}}_\alpha \in \mathbb{R}^{1 \times (d+E)}.
    \end{equation}
    
    If no dimensionality reduction is applied (i.e., the full loci embedding size $L$ is used), the environment embeddings are padded as:
    \begin{equation}
        \mathbf{\bar{e}}_\alpha= \begin{bmatrix} \mathbf{0}_{1 \times L} & \mathbf{e}_\alpha \end{bmatrix}, \quad \mathbf{\bar{e}}_\alpha \in \mathbb{R}^{1 \times (L+E)}.
    \end{equation}
    
    \item A constant feature of $1$ is appended to both the loci and environment embeddings at the rightmost position:
    \begin{equation}
        \tilde{Z}^{(g)} = \begin{bmatrix} \bar{Z}^{(g)} & \mathbf{1}_{L \times 1} \end{bmatrix}, \quad \tilde{Z}^{(g)} \in \mathbb{R}^{L \times (d+E+1)}.
    \end{equation}
    \begin{equation}
        \mathbf{\tilde{e}}_\alpha = \begin{bmatrix} \mathbf{\bar{e}}_\alpha & 1 \end{bmatrix}, \quad \mathbf{\tilde{e}}_\alpha \in \mathbb{R}^{1 \times (d+E+1)}.
    \end{equation}
    
    If no dimensionality reduction is applied, the final representations become:
    \begin{equation}
        \tilde{Z}^{(g)} = \begin{bmatrix} \bar{Z}^{(g)} & \mathbf{1}_{L \times 1} \end{bmatrix}, \quad \tilde{Z}^{(g)} \in \mathbb{R}^{L \times (L+E+1)}.
    \end{equation}
    \begin{equation}
        \mathbf{\tilde{e}}_\alpha = \begin{bmatrix} \mathbf{\bar{e}}_\alpha & 1 \end{bmatrix}, \quad \mathbf{\tilde{e}}_\alpha \in \mathbb{R}^{1 \times (L+E+1)}.
    \end{equation}
\end{itemize}

This alignment ensures that both genotype and environment representations share the same dimensional scale, $d + E + 1$ (or $L+E+1$), during attention computations. This structured encoding allows the model to dynamically infer gene-environment interactions.

\subsection{Training strategy}

To effectively capture genotype-phenotype relationships across multiple environments, the model is trained using mini-batch stochastic gradient descent. We ensure that each mini-batch includes samples from all environments, promoting cross-environment learning and enhancing the model’s ability to generalize across diverse conditions. The training setup follows these principles:

    \textit{Balanced representation of environments:} Each mini-batch includes an equal number of individuals from all environments, preventing any single environment from dominating the training process. This ensures that the model learns environment-specific effects in a balanced manner. Given that each mini-batch contains $N_{\text{ind}, \alpha} = 4$ individuals per environment and there are $E = 18$ environments, the total \texttt{batch\_size} is given as:

    \begin{equation}
        \texttt{batch\_size} = \sum_{\alpha=1}^{E} N_{\text{ind}, \alpha} = \sum_{\alpha=1}^{18} 4 = 72. \label{eq_4seg}
    \end{equation}
This setup prevents the model from developing biases toward specific environments and ensures that predictions remain stable rather than fluctuating unpredictably as the mini-batch changes across training iterations.

   \textit{Cross-talking between environments:} Including individuals from all environments in each mini-batch allows the model to simultaneously process information from different environmental conditions and update its parameters based on a comprehensive view of genotype-phenotype interactions across environments.

Since each mini-batch contains $E$ environments, with $N_{\text{ind}, \alpha}$ individuals per environment, the batch-wise loss function aggregates gradients across environments becomes:

\begin{equation}
    \mathcal{L}_{\text{batch}} = \frac{1}{E} \sum_{\alpha=1}^{E} \mathcal{L}_{\alpha}, \label{eq_loss_multi}
\end{equation}
where $\mathcal{L}_{\alpha}$ is the loss computed for $N_{\text{ind}, \alpha}$ individuals from environment $\alpha$. During backpropagation, the gradient updates are computed jointly:

\begin{equation}
    \theta \leftarrow \theta - \eta \frac{1}{E} \sum_{\alpha=1}^{E} \nabla_\theta \mathcal{L}_{\alpha},
\end{equation}
where: $\theta$ represents the model parameters, $\eta$ is the learning rate, and $\nabla_\theta \mathcal{L}_{\alpha}$ is the gradient contribution from environment $\alpha$. This ensures that individuals from different environments influence shared model parameters, allowing information from one environment to shape predictions in another, even without direct cross-attention. 

The remainder of the training procedure, as well as the validation and testing processes, are identical to those described for the single-environment attention model in Section \ref{sec_app_3}.

\section{Transfer learning \label{sec_app_5}}

In scenarios where data for a specific environment is sparse, the multi-environment attention-based model can leverage correlations across environments to improve predictions in the underrepresented environment. This approach enables the model to infer missing phenotypic values by exploiting shared genetic and environmental structures.

Let there be a total of $E$ environments, each represented as a one-hot vector $\mathbf{e}_\alpha\in \mathbb{R}^{E}$. In our work, there are eight temperatures (see Eq.~(\ref{eq_temps})), so $E=8$. For the synthetic data, we use $d = 30$, and thus $d > E$.  To match the dimensionality of the loci embeddings,  the environment tokens are padded with zeros such that the rightmost $d - E$ values are set to zero. This ensures compatibility within the attention mechanism while maintaining a distinct representation for each environment. The padded environment embedding is defined as:

\begin{equation}
    \mathbf{\bar{e}}_\alpha= \begin{bmatrix} \mathbf{e}_\alpha & \mathbf{0}_{1 \times (d - E)} \end{bmatrix}, \quad \mathbf{\bar{e}}_\alpha \in \mathbb{R}^{1 \times d}.
\end{equation}
A constant feature of $1$ is then appended to maintain compatibility with loci embeddings (Eq.~(\ref{eq_Z_with1})):

\begin{equation}
   \mathbf{\tilde{e}}_\alpha = \begin{bmatrix} \mathbf{\bar{e}}_\alpha& 1 \end{bmatrix}, \quad \mathbf{\tilde{e}}_\alpha \in \mathbb{R}^{1 \times (d+1)}.
\end{equation}
This transformation ensures that both genotype and environment embeddings have the same dimension $d+1$, allowing them to be processed seamlessly within the attention mechanism.

To simulate data sparsity in a single environment, we randomly select one environment—let us call it the environment $\alpha^*$—and introduce missing values. This is done by randomly selecting $N_{\alpha^*} - M_{\alpha^*}$ fitness values within this environment and setting them to \texttt{NaN} (Not a Number).
In computing, \texttt{NaN} is a special placeholder used to represent missing, undefined, or unrepresentable numerical values. Here, $N_{\alpha^*}$ represents the total number of training samples in environment $\alpha^*$, and $M_{\alpha^*}$ denotes the number of samples available to the model during training. The training samples from all other environments remain fully available.

The mini-batch structure is designed to maintain a balanced representation across environments. Each batch consists of four individuals per environment (see Eq.~(\ref{eq_4seg})). However, when computing the loss function (see  Eq.~(\ref{eq_loss_multi})), individuals with NaN fitness values are discarded, ensuring that only available data contribute to parameter updates.

This setup allows the underrepresented environment, $\alpha^*$, to interact with all other environments during every mini-batch processing, facilitating cross-environment learning. 

The remainder of the training procedure, as well as the validation and testing processes, are identical to those described for the single-environment attention model in Section \ref{sec_app_3}.

\section{Series expansions of an attention layer \label{sec_app_6}}

In this section, we derive a series expansion for the output of a single attention layer, considering the case where one-hot encoded loci vectors are fed directly into the layer without any dimensionality reduction. We use the following notations throughout:
\begin{itemize}
    \item $ \mathbf{z}_i $:  Embedding vector for the $i^{\text{th}}$ locus.
    \item $ \mathbf{q}_{i} $:  Query vector for the $i^{\text{th}}$ locus.
    \item $ \mathbf{k}_{i} $:  Key vector for the $i^{\text{th}}$ locus.
    \item $ \mathbf{v}_{i} $:  Value vector for the $i^{\text{th}}$ locus.
    \item $ W^Q $, $ W^K $, $ W^V $: Query, key, and value projection matrices, respectively.
    \item $ C^i $: Coefficients of the $i$-th order interaction.
    \item $ \beta $: Regression coefficients for the attended vectors.
    \item $ x_{i} \in \{+1,-1\} $:  SNP value associated with the $i^{\text{th}}$ locus.
\end{itemize}

We start by considering the scenario where the one-hot encoding of the loci is directly provided to the attention layer without any prior dimensionality reduction.

The $\nu^{\text{th}}$ component of the value vector $ \mathbf{v}_{j} $, denoted by $ v_{j \nu} $, is computed as follows:
\begin{equation}
v_{j \nu} = \sum_{m} z_{j m} \, W^V_{m \nu}.
\end{equation}
Since the one-hot embedding implies that $ z_{j m} = \delta_{jm}\, x_{j} $ (where $ \delta_{jm} $ is the Kronecker delta), this expression simplifies to:
\begin{equation}
v_{j \nu} = \sum_{m} \delta_{jm}\, x_{j} \, W^V_{m \nu} = x_{j}\, W^V_{j \nu}. \label{eq_vjnu}
\end{equation}
Similarly, we obtain $ q_{i m} = x_{i}\, W^Q_{i m} $ and $ k_{j m} = x_{j}\, W^K_{j m} $.

The dot product between the query vector $ \mathbf{q}_{i} $ and the key vector $ \mathbf{k}_{j} $ is given by:
\begin{equation}
\mathbf{q}_{i} \cdot \mathbf{k}_{j} = \sum_{m} q_{i m} \, k_{j m}.
\end{equation}
Given that $ q_{i m} = x_{i}\, W^Q_{i m} $ and $ k_{j m} = x_{j}\, W^K_{j m} $, we obtain:
\begin{equation}
\mathbf{q}_{i} \cdot \mathbf{k}_{j} = \sum_{m} x_{i}\, W^Q_{i m}\, x_{j}\, W^K_{j m} = x_{i}\, x_{j}\, \alpha_{ij},\label{eq_qk_dot}
\end{equation}
where we define
\begin{equation}
\alpha_{ij} = \sum_{m} W^Q_{i m} \, W^K_{j m}.
\end{equation}

The predicted fitness, denoted by $ y_{\text{pred}} $, is given by:
\begin{equation}
y_{\text{pred}} = \sum_{i, \nu}\left[\sum_{j} \frac{e^{\mathbf{q}_{i} \cdot \mathbf{k}_{j}}}{\displaystyle\sum_{m} e^{\mathbf{q}_{i} \cdot \mathbf{k}_{m}}} \, v_{j \nu}\right] \beta_{i \nu} + \beta_0. \label{eq_pred_fit_appendix}
\end{equation}

We expand the softmax term as a Taylor series:
\begin{equation}
\frac{e^{\mathbf{q}_{i} \cdot \mathbf{k}_{j}}}{\displaystyle\sum_{m} e^{\mathbf{q}_{i} \cdot \mathbf{k}_{m}}} = \frac{1 + \mathbf{q}_{i} \cdot \mathbf{k}_{j} + \frac{(\mathbf{q}_{i} \cdot \mathbf{k}_{j})^2}{2!} + \frac{(\mathbf{q}_{i} \cdot \mathbf{k}_{j})^3}{3!} + \cdots}{\displaystyle\sum_{m} \left( 1 + \mathbf{q}_{i} \cdot \mathbf{k}_{m} + \frac{(\mathbf{q}_{i} \cdot \mathbf{k}_{m})^2}{2!} + \frac{(\mathbf{q}_{i} \cdot \mathbf{k}_{m})^3}{3!} + \cdots \right)}.
\end{equation}
Inserting $ \mathbf{q}_{i} \cdot \mathbf{k}_{j} = x_{i}\, x_{j}\, \alpha_{ij} $ from Eq.~(\ref{eq_qk_dot}), we have:
\begin{equation}
\frac{e^{\mathbf{q}_{i} \cdot \mathbf{k}_{j}}}{\displaystyle\sum_{m} e^{\mathbf{q}_{i} \cdot \mathbf{k}_{m}}} = \frac{1 + x_{i}\, x_{j}\, \alpha_{ij} + \frac{\alpha_{ij}^2}{2} + x_{i}\, x_{j}\, \frac{\alpha_{ij}^3}{6} + \cdots}{\displaystyle\sum_{m}\left(1 + x_{i}\, x_{m}\, \alpha_{im} + \frac{\alpha_{im}^2}{2} + x_{i}\, x_{m}\, \frac{\alpha_{im}^3}{6} + \cdots\right)}.
\end{equation}
We can express the series in terms of hyperbolic functions by recognizing that:
\begin{equation}
1 + x_{i}\, x_{j}\, \alpha_{ij} + \frac{\alpha_{ij}^2}{2} + x_{i}\, x_{j}\, \frac{\alpha_{ij}^3}{6} + \cdots = \frac{e^{\alpha_{ij}} + e^{-\alpha_{ij}}}{2} + x_{i}\, x_{j}\, \frac{e^{\alpha_{ij}} - e^{-\alpha_{ij}}}{2}.
\end{equation}
Thus, the expression becomes:
\begin{equation}
\frac{e^{\mathbf{q}_{i} \cdot \mathbf{k}_{j}}}{\displaystyle\sum_{m} e^{\mathbf{q}_{i} \cdot \mathbf{k}_{m}}} = \frac{\beta_{ij}^{+} + x_{i}\, x_{j}\, \beta_{ij}^{-}}{\displaystyle\sum_{m}\left(\beta_{im}^{+} + x_{i}\, x_{m}\, \beta_{im}^{-}\right)},
\end{equation}
where we have defined
\begin{equation}
\beta_{ij}^{\pm} = \frac{e^{\alpha_{ij}} \pm e^{-\alpha_{ij}}}{2}.
\end{equation}
Let
\begin{equation}
g_{i}^{+} = \sum_{m} \beta_{im}^{+} \quad \text{and} \quad h_{ij}^{\pm} = \frac{\beta_{ij}^{\pm}}{g_{i}^{+}}.
\end{equation}
With these definitions, the softmax expression simplifies to:
\begin{equation}
\frac{e^{\mathbf{q}_{i} \cdot \mathbf{k}_{j}}}{\displaystyle \sum_{m} e^{\mathbf{q}_{i} \cdot \mathbf{k}_{m}}} = \frac{h_{ij}^{+} + x_{i}\, x_{j}\, h_{ij}^{-}}{1 + x_{i} \displaystyle\sum_{m} x_{m}\, h_{im}^{-}}.
\end{equation}

Now, substituting the above and the expression for $ v_{j \nu} $ from Eq.~(\ref{eq_vjnu}) into Eq.~(\ref{eq_pred_fit_appendix}), we obtain:
\begin{equation}
y_{\text{pred}} = \sum_{i,j,\nu} \frac{x_{j}\, h_{ij}^{+} + x_{i}\, h_{ij}^{-}}{1 + x_{i} \displaystyle\sum_{m} x_{m}\, h_{im}^{-}}\, W^V_{j \nu}\, \beta_{i \nu} + \beta_0.
\end{equation}
Expanding the denominator as a series, we have:
\begin{equation}
\frac{1}{1 + x_{i} \displaystyle\sum_{m} x_{m}\, h_{im}^{-}} \approx 1 - x_{i} \sum_{m} x_{m}\, h_{im}^{-} + \sum_{m,m'} x_{m}\, x_{m'}\, h_{im}^{-}\, h_{im'}^{-} - x_{i} \sum_{m,m',m''} x_{m}\, x_{m'}\, x_{m''}\, h_{im}^{-}\, h_{im'}^{-}\, h_{im''}^{-} + \cdots.
\end{equation}
Thus, the predicted fitness can be expressed as:
\begin{equation}
y_{\text{pred}} = \sum_{i,j,\nu} W^V_{j \nu}\, \beta_{i \nu}\, \left(x_{j}\, h_{ij}^{+} + x_{i}\, h_{ij}^{-}\right) \left(1 - x_{i} \sum_{m} x_{m}\, h_{im}^{-} + \sum_{m,m'} x_{m}\, x_{m'}\, h_{im}^{-}\, h_{im'}^{-} - \cdots \right) + \beta_0.
\end{equation}
This expansion can be rearranged to show:
\begin{equation}
y_{\text{pred}} = C^0 + \sum_{i} C_{i}^{1}\, x_{i} + \sum_{i,j,k} C_{ijk}^{3}\, x_{i}\, x_{j}\, x_{k} + \cdots.
\end{equation}
Notice that this series does not include explicit pairwise interactions between loci (i.e., terms like $x_{i}\, x_{j}$ do not appear on their own). 

\subsection*{Changing embedding to incorporate pairwise interactions}

We consider a simple modification to the embedding—appending an additional element $1$ at the end of the one-hot vectors. With this change, 
$$
z_{i \nu}=\delta_{i \nu}\, x_{i}+\delta_{L+1, \nu},
$$ 
so that the new embedding is $(L+1)$-dimensional. The value vector is then given as:
\begin{align}
v_{j \nu} & =\sum_{m} z_{j m}\, W^V_{m \nu} \nonumber\\
& =\sum_{m}\left(\delta_{j m}\, x_{j}+\delta_{L+1, m}\right)\, W^V_{m \nu} \nonumber\\
& =x_{j}\, W^V_{j \nu}+W^V_{L+1, \nu}\label{eq_vjn_2} ,
\end{align}
and the dot product $ \mathbf{q}_{i} \cdot \mathbf{k}_{j} $ can be obtained as
\begin{align}
\mathbf{q}_{i} \cdot \mathbf{k}_{j} & =\sum_{m} q_{i m}\, k_{j m} \nonumber\\
& =\sum_{m}\left(x_{i}\, W^Q_{i m}+W^Q_{L+1, m}\right)\left(x_{j}\, W^K_{j m}+W^K_{L+1, m}\right) \nonumber\\
& =x_{i}\, x_{j}\, \alpha_{ij}+x_{i}\, \beta_{i}+x_{j}\, \gamma_{j}+s\label{eq_q.k_2} ,
\end{align}
where
\begin{equation}
\begin{aligned}
\alpha_{ij} &= \sum_{m} W^Q_{i m}\, W^K_{j m}, \quad \eta_{i}=\sum_{m} W^Q_{i m}\, W^K_{L+1, m},\\
\gamma_{j} &= \sum_{m} W^Q_{L+1, m}\, W^K_{j m}, \quad s=\sum_{m} W^Q_{L+1, m}\, W^K_{L+1, m}.
\end{aligned}
\end{equation}

Without the softmax, the fitness $y_{\text{pred}}$ is given by:
\begin{equation}
y_{\text{pred}}=\sum_{i,\nu}\left[\sum_{j} \left(\mathbf{q}_{i} \cdot \mathbf{k}_{j}\right)\, v_{j \nu}\right]\, \beta_{i \nu}+\beta_0\label{eq_fit_2} .
\end{equation}
Substituting Eqs.~(\ref{eq_vjn_2}) and (\ref{eq_q.k_2}) into Eq.~(\ref{eq_fit_2}), we obtain:
\begin{align}
y_{\text{pred}}= &  \sum_{i,j,\nu}\left(x_{i}\, x_{j}\, \alpha_{ij}+x_{i}\, \eta_{i}+x_{j}\, \gamma_{j}+s\right)
\left(x_{j}\, W^V_{j \nu}+W^V_{L+1, \nu}\right)\, \beta_{i \nu}+\beta_0 \nonumber\\
=& \; C^{0}+\sum_{i} C_{i}^{1}\, x_{i}+\sum_{i,j} C_{ij}^{2}\, x_{i}\, x_{j} ,
\end{align}
where
\begin{align*}
C^{0} &= \beta_0 +\sum_{i,j,\nu} \beta_{i \nu}\left(\gamma_{j}\, W^V_{j \nu}+s\, W^V_{L+1, \nu}\right),\\
C_{i}^{1} &= \sum_{\nu,j}\left\{\beta_{i \nu}\left(\alpha_{ij}\, W^V_{j \nu}+\eta_{i}\, W^V_{L+1, \nu}\right)+\beta_{j \nu}\left(\gamma_{i}\, W^V_{L+1, \nu}+s\, W^V_{i \nu}\right)\right\},\\
C_{ij}^{2} &= \sum_{\nu} \beta_{i \nu}\left(\alpha_{ij}\, W^V_{L+1, \nu}+\eta_{i}\, W^V_{j \nu}\right).
\end{align*}

Even without the softmax, we observe an explicit pairwise interaction term. With the softmax applied, higher-order terms naturally emerge. This motivates us to always append a constant value of 1 to the end of all embedding vectors.

\begin{figure}
     \centering
         \includegraphics[width=\textwidth]{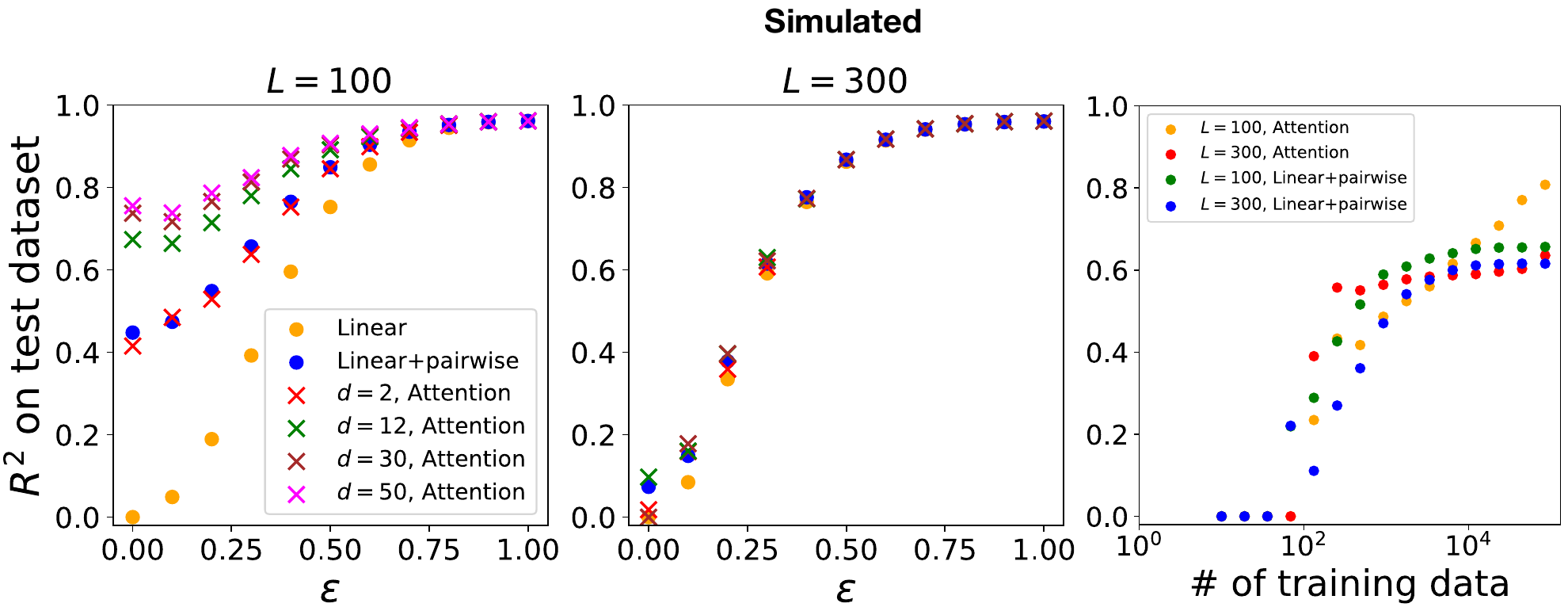}        
 \caption{\textbf{Performance comparison of models for exponentially distributed coefficients $\beta$ with varying embedding dimensions $d$ and number of loci $L$.} The left and center panels show $R^2$ values on the test dataset for $L = 100$ and $L = 300$, respectively, across different epistasis strengths $\epsilon$.  The right panel illustrates $R^2$ performance at $\epsilon = 0.3$ as a function of training dataset size, using an embedding dimension of $d = 30$ for the attention-based models. \label{fig_app_2}} 
\end{figure}
\clearpage

\begin{figure}
     \centering
         \includegraphics[width=\textwidth]{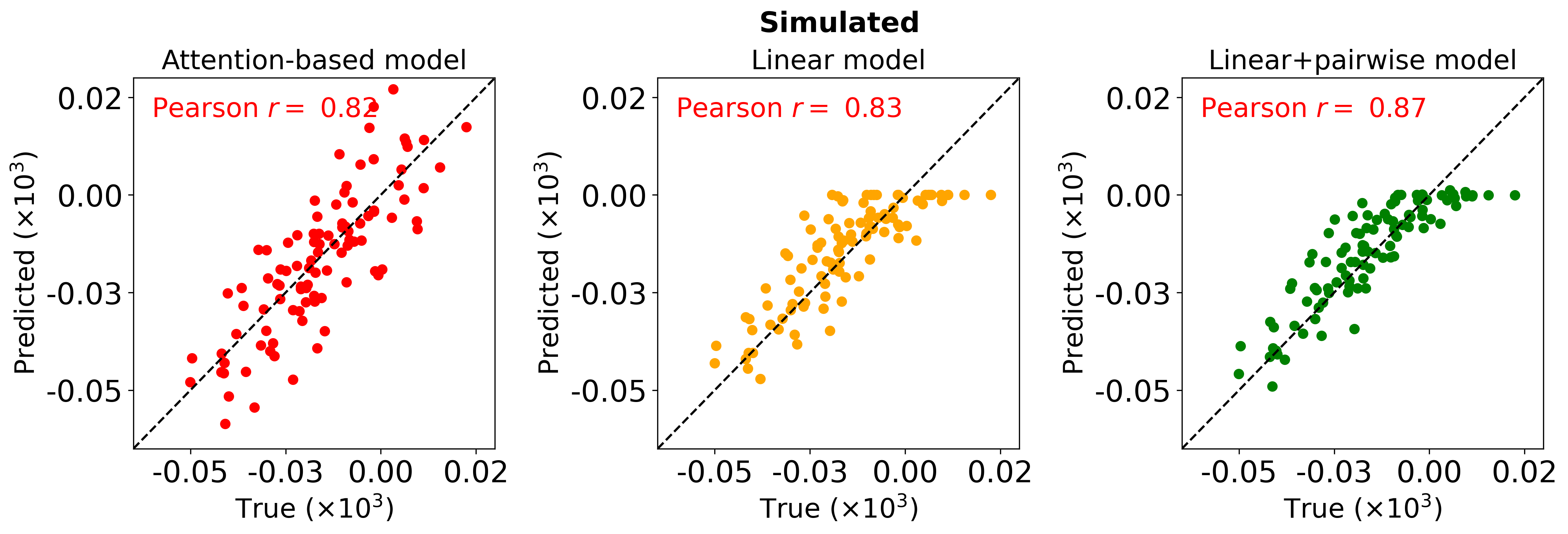}        
    \caption{\textbf{Comparison of predicted effect sizes for Gaussian-distributed coefficients with $L = 100$ and $\epsilon = 0.3$.} We compare the predicted effect sizes, averaged over genotype backgrounds, on the test dataset across three models:  attention-based model with $d = 30$ (left),  linear model (middle), and  linear+pairwise model (right). Each plot shows the correlation between the predicted and true effect sizes, with the respective Pearson correlation coefficients indicated.}
        \label{fig_app_3}
\end{figure}

\begin{figure}
     \centering
         \includegraphics[width=\textwidth]{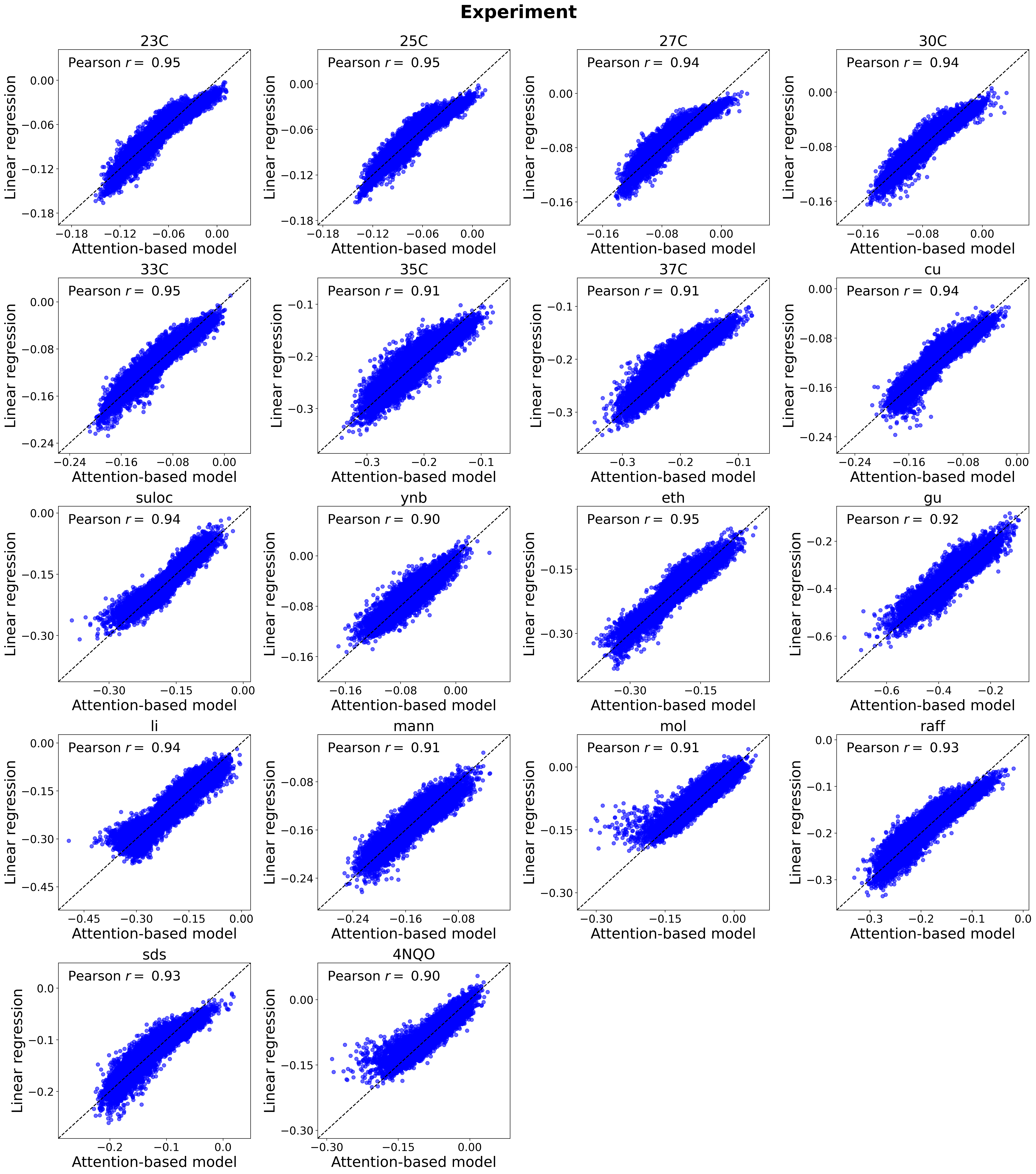}        
   \caption{\textbf{Comparison of fitness predictions from linear regression and a single-environment attention-based model across multiple environments.}
    Each subplot corresponds to one of eighteen different media conditions (indicated in the subplot titles) and shows the predicted fitness values for all test-set individuals.
    The $y$-axis represents the linear regression predictions, while the $x$-axis shows those from the attention-based model.
    Pearson correlation coefficients are shown in each subplot.
}\label{fig_app_4}
\end{figure}

\begin{figure}
     \centering
         \includegraphics[width=\textwidth]{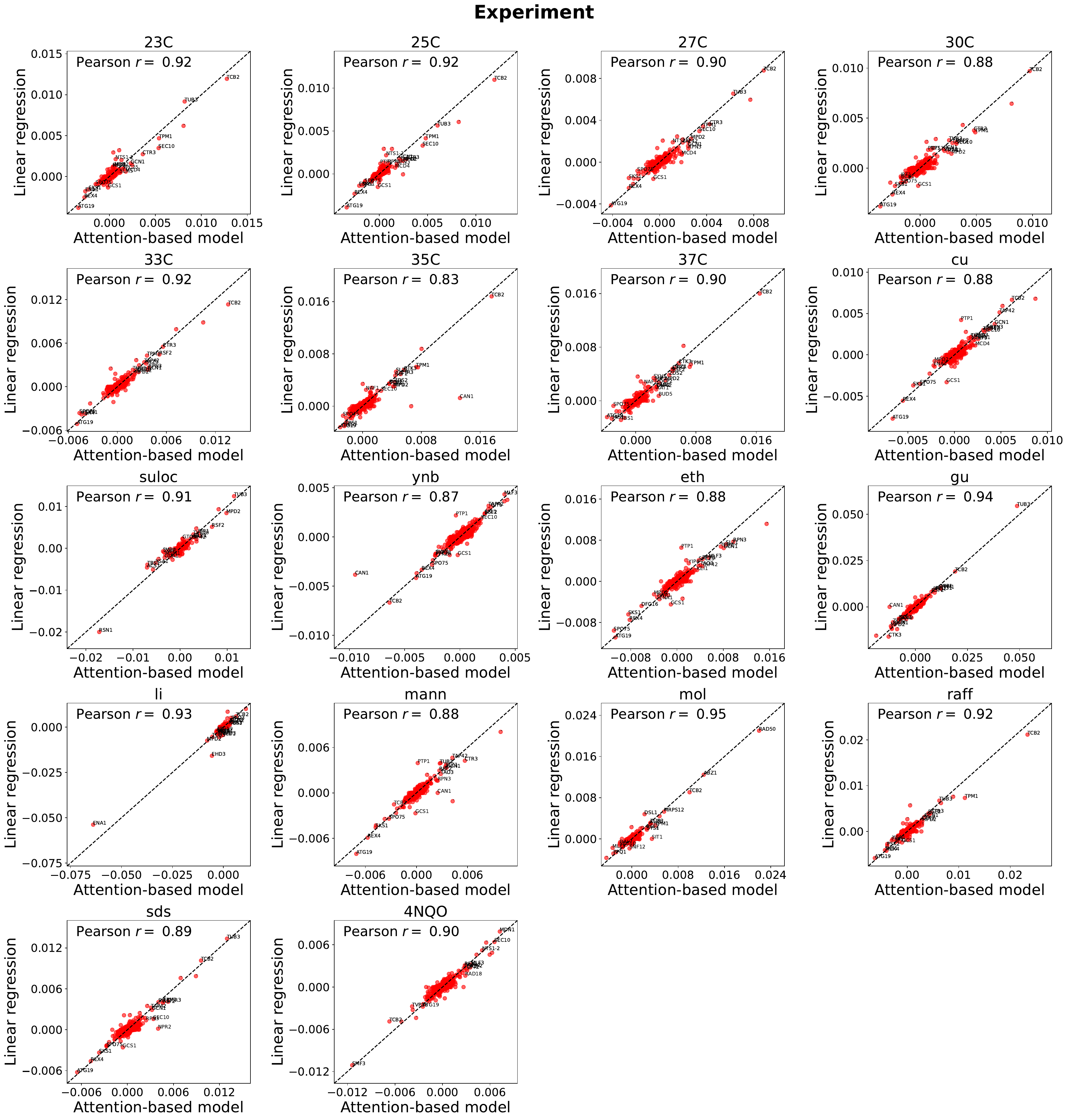}        
    \caption{\textbf{Comparison of effect sizes predicted by linear regression and a single-environment attention-based model under different conditions.} 
    Each subplot corresponds to a different environmental condition and compares effect sizes predicted by the two models. 
    The loci with the 20 largest effect sizes are annotated. Pearson correlation coefficients are shown in each subplot.}\label{fig_app_5}
\end{figure}

\begin{figure}
     \centering
         \includegraphics[width=\textwidth]{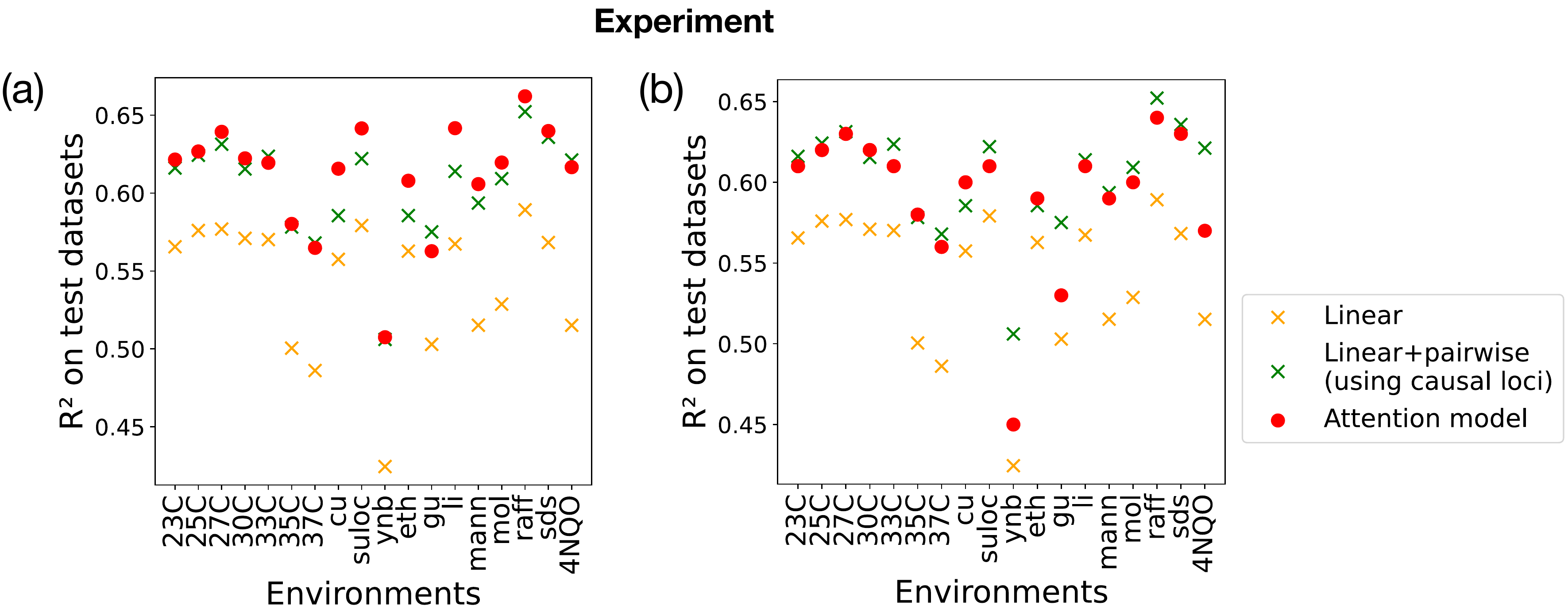}    
     \caption{\textbf{Performance comparison of attention-based architectures with one-hot embeddings for loci and environment.} 
    The $R^2$ values on test datasets are shown for the linear model, the linear+pairwise model (using causal loci), and the attention-based model across various environments. 
    (a) Single-environment attention architecture. (b) Multi-environment attention architecture. Both architectures use two attention layers.}
        \label{fig_app_6}
\end{figure}

\begin{figure}
     \centering
         \includegraphics[width=\textwidth]{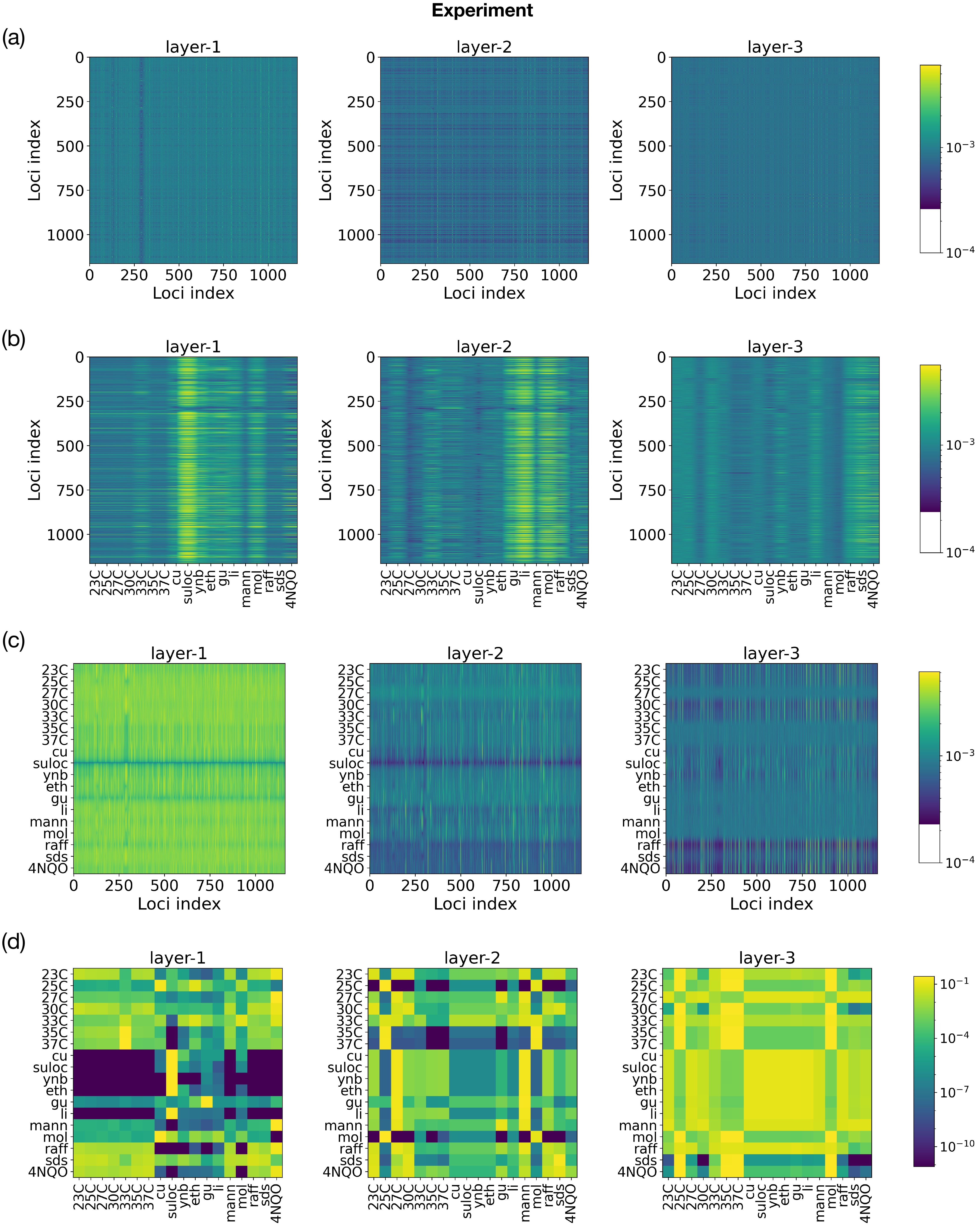}        
  \caption{
\textbf{Heatmaps of attention scores across three successive layers in the multi-environment architecture with $d=12$.} 
    The scores are computed using Eq.~(\ref{eq_atten_score}) and averaged across all test segregants. 
    The horizontal axis represents key vectors, and the vertical axis represents query vectors.  
    Panels (a)--(d) show locus--locus, environment--locus, environment--locus, and environment--environment attention maps, respectively.   Note that the key and query vectors for environments are independent of the individuals.
}\label{fig_app_7}
\end{figure}

\end{document}